\documentclass[11pt,showpacs,nofootinbib,showkeys]{revtex4}
\usepackage{graphicx,color,amsmath,amssymb}
\usepackage{epstopdf}
\usepackage{mathrsfs}

\newcommand{\too}{\longrightarrow}
\newcommand{\oto}{\leftrightarrow}
\newcommand{\Cal}[1]{{\cal #1}}
\newcommand{\be}{\begin{equation}}
\newcommand{\ee}{\end{equation}}
\newcommand{\bea}{\begin{eqnarray}}
\newcommand{\eea}{\end{eqnarray}}
\newcommand{\p}{\partial}
\newcommand{\upa}{\uparrow}
\newcommand{\down}{\downarrow}
\newcommand{\ket}[1]{\left| #1 \right\rangle}
\newcommand{\bra}[1]{\left\langle #1 \right|}
\newcommand{\ind}[1]{\begin{footnotesize}\mbox{#1}\end{footnotesize}}
\newcommand{\acom}[2]{\left\{#1,#2\right\}}
\newcommand{\com}[2]{\left[#1,#2\right]}
\newcommand{\esp}{\,,\hspace{0.6cm}}
\newcommand{\vm}[1]{\left\langle #1 \right\rangle}
\newcommand{\dr}[1]{\mbox{#1}}
\newcommand{\fr}[1]{\mathfrak{#1}}
\newcommand{\frg}{\mathfrak{g}}
\newcommand{\f}{\textsc{f}}
\newcommand{\R}{\textsc{r}}
\newcommand{\ens}[2]{\left\{#1\right\}_{#2}}
\renewcommand{\L}{\textsc{l}}
\renewcommand{\P}{\textsc{p}}
\newcommand{\I}{$I\ $}
\newcommand{\II}{${I \! \! I\ }$}
\newcommand{\III}{${I \! \! I\!\! I\ }$}
\renewcommand{\tilde}[1]{\widetilde{#1}}
\newcommand{\nl}{\nonumber\\}
\newcommand{\idy}{{1\!\!1}}
\newcommand{\re}{\Re\mathfrak{e}}
\newcommand{\im}{\Im\mathfrak{m}}
\newcommand{\hc}{\mbox{H.c.}}
\newcommand{\J}{{\cal J}} % WZNW current
\newcommand{\tJ}{\widetilde{\J}}
\newcommand{\du}{\mathscr D} % set of dualities
\newcommand{\M}{\mathscr M} % model (defined by bare couplings)
\newcommand{\Z}{\mathbb{Z}}
\newcommand{\tr}{\mbox{Tr}}
\newcommand{\downto}{\leadsto}
\newcommand{\coxeter}{g^\vee}
\newcommand{\gwz}{{\texttt g}}
\newcommand{\low}{{\vphantom{\dagger}}}
\newcommand{\DSE}{\textsc{dse}}

\newcommand{\drawbunchesmall}{\begin{tabular}{rcl}
\scriptsize $p$ & $\big\{$ &
\begin{tabular}{c}
   \hline\hphantom{aaaaaaaaa}\vspace*{-0.59cm}\\
   \hline\vspace*{-0.59cm}\\
   \hline \vspace*{-0.92cm}\\
   $\cdot$ \vspace*{-0.56cm}\\
   $\cdot$ \vspace*{-0.56cm}\\
   $\cdot$ \vspace*{-0.21cm}\\
    \hline
\end{tabular}
\vspace*{-0.12cm}\\
\scriptsize $q=N-p$ & $\big\{$ &
\begin{tabular}{c}
   \hline\hphantom{aaaaaaaaa}\vspace*{-0.59cm}\\
   \hline \vspace*{-0.92cm}\\
   $\cdot$ \vspace*{-0.56cm}\\
   $\cdot$ \vspace*{-0.56cm}\\
   $\cdot$ \vspace*{-0.21cm}\\
   \hline
\end{tabular}
\end{tabular}
}

\newcommand{\drawsymbreak}{
\begin{tabular}{ccccc}
&& $\;$SO(3)$\times$SO(3) $\;$ &&
\vspace*{-0.5cm}\\
&  $\diagup$ && $\diagdown$  &
\vspace*{-0.1cm}\\
 SO(3)$\times$U(1)$\times\Z_2$ $\;\;$&
\begin{tabular}{c} \hphantom{$\diagup$}\vspace*{-0.85cm}\\\hline\end{tabular} & SO(4)$\times
$SO(2) &
\begin{tabular}{c} \hphantom{$\diagup$}\vspace*{-0.85cm}\\\hline\end{tabular} &
 \quad SU(4)
\vspace*{-0.23cm}\\
&  $\diagdown$ && $\diagup$  &
\vspace*{-0.4cm}\\
&& SO(5)$\times$$\Z_2$ &&
\end{tabular}
}
\begin{document}

\title{Duality approach to one-dimensional
degenerate electronic systems}
\author{E. Boulat}
\email{edouard.boulat@univ-paris-diderot.fr}
\affiliation{Laboratoire Mat\'eriaux et Ph\'enom\`enes Quantiques,
CNRS, UMR 7162,
Universit\'e Paris  Diderot, 
B\^at. Condorcet, 75205 Paris Cedex 13, France}
\author{P. Azaria}
\affiliation{
Laboratoire de Physique Th\'eorique des Liquides,
Universit\'e Pierre et Marie Curie, 4 Place Jussieu,
75256 Paris Cedex 05, France}
\author{P. Lecheminant}
\affiliation{Laboratoire de Physique Th\'eorique et
Mod\'elisation, CNRS UMR 8089,
Universit\'e de Cergy-Pontoise, Site de Saint-Martin,
F-95000 Cergy-Pontoise, France}

%%%%%%%%%%%%%%%%%%%%%%%%%%%%%%%%%%%%%%%%%%%%%
\begin{abstract}
We investigate the possible
classification of zero-temperature spin-gapped phases of multicomponent
electronic systems in one spatial dimension.
At the heart of our analysis is the existence of
non-perturbative duality symmetries which
emerge within a low-energy description.
These dualities
fall into a finite number of classes that can
be listed and
depend only on the algebraic properties of the symmetries of the
system: its physical symmetry group and
the maximal continuous symmetry group of the interaction.
We further characterize possible competing orders
associated to the dualities
and discuss the nature of the quantum phase transitions
between them.
Finally, as an illustration, the duality approach is applied to
the description of the phases of
two-leg electronic ladders at incommensurate filling.
\end{abstract}
%%%%%%%%%%%%%%%%%%%%%%%%%%%%%%%%%%%%%%%%%%%%%

\maketitle

%%%%%%%%%%%%%%%%%%%%%%%%%%%%%%%%%%%%%%%%%%%%%

\section{Introduction}

Conformal Field Theory (CFT) is a powerful approach which
determines the low-energy properties of one-dimensional (1D)
quantum systems or two-dimensional
classical systems close to criticality \cite{bpz,difrancesco97}.
In particular, it allows for a complete classification
of 1D gapless phases and quantum critical points with Lorentz
invariance  \cite{affleck85}.
The low-energy gapless spectrum of the corresponding 1D lattice model with
continuous symmetry, like the spin-1/2 Heisenberg chain,
is described in terms of representations of certain
Kac-Moody (KM) current algebra \cite{kz}. This affine symmetry
fixes the operator content of the system and the 
scaling dimensions of the operators are in
turn determined by the conformal invariance
of the underlying Wess-Zumino-Novikov-Witten (WZNW) model
built from the currents of the affine symmetry \cite{difrancesco97}.
The 1D gapless phases are then labelled
by the central charge $c$ of the corresponding WZNW CFT.
The simplest example of this description is the Luttinger liquid universality
class  which accounts for the low-energy properties
of 1D metals \cite{haldane,bookboso,giamarchi}.
Such a CFT has a U(1) affine KM symmetry with central charge $c=1$.
The critical theory of half-integer Heisenberg spin chains corresponds
to the SU(2)$_1$ WZNW whereas the family of integrable spin chains of
arbitrary spins $S$ admits for a  critical theory the SU(2)$_{2S}$ WZNW
model with central
charge $c = 3S/(S+1)$ \cite{affleckspinchain}.

In sharp contrast to gapless systems,
a general classification of all possible
zero temperature 1D gapped phases
is still lacking despite numerous works in the past two decades.
To the best of our knowledge, there exists no general symmetry based arguments
that allow for a quantitative description of 1D quantum problems
with a spectral gap.
A natural starting approach is to identify the massive
phases that occur in the vicinity of quantum critical
points or close to gapless phases which are described
by some known CFTs.
Very special perturbations could be singled out
which correspond to massive integrable deformations of
the original CFT \cite{zamolo}.
The integrability provides, in turn,
a non-perturbative basis to construct the massive
quasiparticles defining the low-energy spectrum of
the fully gapped phase.
The stability of these excitations can then
be investigated perturbatively with respect
to other non-integrable generic perturbations \cite{mussardo}
and it will define the extension of the massive phase.
A second  more phenomenological approach relies directly on orders by
considering very special ones and the use of
extended symmetries which unify them.
A typical example is the SO(5) theory
for the competition between $d$-wave superconductivity and
antiferromagnetism, where the order parameters of $d$-wave
superconductivity and antiferromagnetism are combined
to form a unified order parameter quintet \cite{zhang}.
Several different competing orders may be transformed into each
other under this extended global symmetry.

In this work, we will combine these two approaches
to describe spin-gapped phases for a large class of 1D
fermionic models where the gap is exponentially small in the
weak coupling regime.
More precisely, we will be concerned with
weakly interacting $N$-component fermionic
models such as $n$-leg electronic
spin-1/2 ladders ($N = 2 n$) \cite{bookboso,giamarchi} or
one-dimensional fermionic cold atoms with
hyperfine spin $F$ ($N = 2 F + 1$) \cite{lecheminant05,lecheminant05b}.
In absence of interactions,
those systems display a large degeneracy, with a
corresponding large
symmetry group mixing microscopic degrees of
freedom, to be called G in the following
\footnote{This definition of the group G as the
microscopic degeneracy group of the
non-interacting theory, G$^{\ind{micro}}$, is not fully
correct. A more rigorous definition of the group G will
be given in section \ref{sectionloweffham}: it is the
maximal symmetry group of
the \emph{interacting} sector of the
theory, which is of course closely related to G$^{\ind{micro}}$.}.
For the aforementioned models,
the group G will be SU($N$) or SO($2N$), depending on the filling.
This degeneracy allows for many possible competing orders.
The low-energy properties of the non-interacting system are
governed by a CFT built on the group G: the G$_k$ WZNW CFT.
For the applications to coupled fermionic chains,
the level of this CFT is $k=1$, but most of our
analysis can be carried out in the general $k$ case.
Interactions  generically break the symmetry G down to a
subgroup H which is the ``true'' physical symmetry of the problem.
The degeneracy lifting which arises from the interactions
results in general in the opening of a spectral gap, which is
accompanied by the stabilization of one of the
possible orders. Questions we will address
in the following are: what kind of
competing orders can be stabilized? How do they
depend on the algebraic properties of the groups involved in the
symmetry breaking
scheme G $\downto$ H?
Can these orders be related between themselves? And
what is the nature of the (zero-temperature) quantum phase transitions?

In this paper, we will investigate these questions by means of
a low-energy approach where the symmetry breaking scheme is
described by  a $G_k$ WZNW model perturbed by
\emph{marginal} current-current interactions with H invariance.
The main result of this work is the fact that
this general class of models possesses
hidden non-trivial duality symmetries which are
associated with
the underlying competing orders. These dualities are emergent,
and exact, in the continuum limit.
They can be viewed as the naive
generalization for continuous symmetry groups of the
well-known Kramers-Wannier (KW) duality symmetry
of the 1D quantum Ising model.
Some duality symmetries are not new and have already
been found in specific models such as the two-leg electronic spin-1/2
ladders at half-filling \cite{balents98,momoi03} or away from half-filling
\cite{schulz,lee,controzzi}, in two-leg
spin ladder with four-spin exchange
interactions \cite{momoi4spin,totsuka}, and also in 1D
multicomponent cold fermions \cite{lecheminant05,lecheminant05b}.
Most of them have been revealed within a one-loop renormalization group (RG)
approach of the interactions with the emergence of a dynamical symmetry
enlargement (DSE) \cite{balents98}.
Such DSE corresponds to a situation where a Hamiltonian is attracted
under RG flow to a manifold possessing a higher symmetry than
the original bare theory.
The main interest of this DSE scenario stems from the fact
that the isotropic RG ray is usually described by an integrable field
theory. The integrability, in turn, leads to the description of
the low-lying excitations and the determination
of the physical properties of the corresponding
spin-gapped phases \cite{balents98,saleur,esslerkonik,essler}.
However, as it will be shown in the following,
the duality symmetries are not restricted to a
two-leg ladder or specific to the one-loop RG calculation
but correspond to a general powerful non-perturbative approach in 1D.
In particular, they can be classified and depend only on the
algebraic properties of G
and of
the actual symmetry H of the lattice model.
As it will be shown,
the resulting list of dualities turns out to be in correspondence with
the  classification of $\Z_2$ graduations of semi-simple
Lie algebra.
The interplay between these dualities and the DSE phenomenon
will also be clarified in this work.
In simple words, the dualities correspond to the different possible
DSEs of the problem.
The existence of these dualities symmetries enables one to
relate different possible competing orders and to shed light on
the global structure of the
phase diagram of weakly-interacting fermionic models.
Of course, our work does not solve the general problem of classification
of zero-temperature 1D gapped phases but, at least,
the duality approach provides a classification of
a subset of gapped phases from symmetry arguments.
Moreover, the nature
of the quantum phase transitions between these orders
can also be determined, within our approach, by
investigating the physics along the self-dual manifolds.

The rest of the paper is organized as follows.
The general low-energy effective theory, that we will study in this work,
is presented in section II.
Section III contains our most important results with
the definition and characterization of the duality symmetries.
These hidden discrete symmetries are introduced from
two different complementary approaches.
A first definition stems from the covariance
of the low-energy effective Hamiltonian.
The characterization and classification of these dualities
are then deduced in section III A.
A second definition is presented in section III B in light of
the DSE phenomenon
where duality symmetries are viewed as exact isometries
of the RG beta-function.
The nature of quantum phase transitions between different
orders is then investigated.
In section IV, the duality approach is applied to
weakly-interacting multicomponent fermionic models
for incommensurate filling.
The special example of a generalized two-leg ladder is then considered
in section V.
Finally, we conclude in section VI
and some technical details on the duality approach
are presented in an appendix.
%%%%%%%%%%%%%%%%%%%%%%%%%%%%%%%%%%%%%%%%%%%%%
%%%%%%%%%%%%%%%%%%%%%%%%%%%%%%%%%%%%%%%%%%%%%

\section{General low-energy effective Hamiltonian}
\label{sectionloweffham}

Let us consider $n$ chains of weakly-interacting
lattice spin-1/2 fermions $c_{i,a}$ $a=1,..,N$ where $N=2n$ is
the total number of internal degrees of freedom of the problem.
The generic model, that we have in mind in this paper, is described by a
lattice Hamiltonian of the Hubbard type with contact interactions:
\begin{equation}
H = - t \sum_{i}\sum_{a} \; \left(
c^\dagger_{i,a}c^{\vphantom{\dagger}}_{i+1,a} + \hc \right)
+\sum_{i}\sum_{a_j}U_{\{a_j\}}\;c^\dagger_{i,a_1}c^\dagger_{i,a_2}
c^{\vphantom{\dagger}}_{i,a_3}c^{\vphantom{\dagger}}_{i,a_4} ,
\label{HamLatt}
\end{equation}
where $t$ is the hopping term
and the independent couplings
(on-site couplings $U_{\{a_j\}}$) depend on
the precise model under consideration, that can be
specified by fixing its physical symmetry group H.
In this respect, many other terms could be added to model (\ref{HamLatt}),
without affecting the following discussion which is
mostly based on symmetry considerations. For example, more general
short-range interaction terms could be considered
as long as they respect the symmetry H of the model.
More concrete examples will be discussed in section V.

In the continuum limit, the non-interacting spectrum around the
two Fermi points $\pm k_\f$ is linearized and gives rise
to left- and right-moving Dirac fermions
$\Psi_{a \L(\R)}$ \cite{bookboso,giamarchi}.
The left (right)-moving fermions are holomorphic (antiholomorphic)
fields of the complex coordinate $z= v_{\f}\tau + ix$ ($\tau$ being
the imaginary time and $v_{\f}$ is the Fermi velocity):
$\Psi_{a \L}(\bar z), \Psi_{a \R}(z)$.
They are normalized by the following operator product expansion (OPE):
\begin{eqnarray}
\Psi_{a \R}^{\dagger} \left(\bar z \right)
\Psi_{b \R} \left(\bar \omega \right) &\sim&
\frac{\delta_{a b}}{2 \pi \left( \bar z - \bar \omega \right)}
\nonumber \\
\Psi_{a \L}^{\dagger} \left(z \right) \Psi_{b \L} \left( \omega \right) &\sim&
\frac{\delta_{a b}}{2 \pi \left(z - \omega \right)} .
\label{OPEfer}
\end{eqnarray}
In the weak-coupling regime, the effect of the interactions is described by
the following general low-energy effective Hamiltonian density:
\begin{eqnarray}
\Cal H = \Cal H_0 + \Cal H_{\ind{int}}
=  -i v_\f \left( \Psi_{a \R}^{\dagger} \partial_x \Psi_{a \R}
- \Psi_{a \L}^{\dagger} \partial_x \Psi_{a \L} \right)
+ \; g_{\alpha} \;{\J}_{\R}^A \; d^{\alpha}_{AB} \;{\J}_{\L}^B,
\label{hamiltonian}
\end{eqnarray}
where velocity renormalization terms and interactions
with the same chirality will be neglected in this work
and $g_{\alpha}$ accounts for the coupling
constants of the fermionic models.
In the following, repeated indices are summed
throughout this paper unless stated otherwise.
The continuous symmetry of the non-interacting
Hamiltonian of Eq. (\ref{HamLatt})
is U($N$) $=$ U(1) $\times$ SU($N$)  where U(1)  is the
usual global charge symmetry
($c_{i,a} \rightarrow
e^{i \phi} c_{i,a}$) and
the SU($N$) group describes the continuous
symmetry of the remaining degrees of freedom, dubbed in the following
``spin'' sector for simplicity.
This basis, which singles out the
charge degrees of freedom, is natural
for incommensurate filling since
a spin-charge separation is expected in the continuum
limit so that
model (\ref{hamiltonian})  decomposes into two commuting
pieces: ${\cal H} = {\cal H}_c + {\cal H}_s$.
In this case, the charge degrees of freedom
display metallic properties in the
Luttinger liquid universality class  \cite{haldane,bookboso,giamarchi}.
The Hamiltonian in the spin sector depends only on the currents
${\J}_{\L(\R)}^A =  2\pi \Psi_{a \L(\R)}^{\dagger}
T^A_{ab}  \Psi_{b \L(\R)}^\low$,
$T^A$ ($A=1,..,N^2 -1$) being the generators of SU($N$)
in the fundamental representation (we choose
the normalization $\tr(T^AT^B)=\delta^{AB}$)
and reads as follows:
\begin{equation}
{\cal H}_s = \frac{v_{\f}}{4\pi\left(N+1\right)}
\left({\cal J}_{\R}^A {\cal J}_{\R}^A + {\cal J}_{\L}^A {\cal J}_{\L}^A
\right)
+ \; g_{\alpha} \;{\J}_{\R}^A \; d^{\alpha}_{AB} \;{\J}_{\L}^B ,
\label{hamspinsector}
\end{equation}
where, in the following, normal ordering
is always implied for the non-interacting term.
Model (\ref{hamspinsector}) is nothing but the
SU($N$)$_1$ WZNW model perturbed by a current-current
interaction.

In contrast, there is no spin-charge
separation in the half-filled case and
the charge degrees of freedom cannot
be disentangled from the
spin ones due to an umklapp process.
A good starting point is then
to consider the maximal global continuous symmetry
of the non-interacting lattice Hamiltonian (\ref{HamLatt})
i.e. G $=$ SO($2N$).
Physically, the appearance of the group SO($2N$)
can be motivated as follows:
due to particle-hole symmetry present at half-filling,
the $2N$ local states  $\big\{c^\dagger_{i,a}\ket{GS} ,
c^{\vphantom{\dagger}}_{i,a}\ket{GS}\big\}$ can play a symmetric role,
where $\ket{GS}$ is a particle-hole symmetric
many-body ground state.
In the continuum limit, the SO($2N$) symmetry
can be revealed
by introducing $2N$ real (Majorana) fermions
from the $N$ Dirac ones:
$\Psi_{a \L(\R)} = (
\xi_{a } + i \; \xi_{a + N})_{\L(\R)}/\sqrt{2}$.
The non-interacting Hamiltonian of model (\ref{hamiltonian})
then takes a
manifestly SO($2N$) invariant form:
$\Cal H_0=-i v_\f \left(\xi_{ a \R}
\partial_x\xi_{ a \R}
- \xi_{a \L}\partial_x\xi_{a \L}\right)/2$.
Within this representation, the currents ${\J}_{\L(\R)}^A$
in Eq. (\ref{hamiltonian})
express as fermionic bilinears:
${\J}_{\L(\R)}^A = i 2 \pi
\xi_{a \L(\R)} \xi_{b \L(\R)}$, $A=(a,b), 1 \le a < b \le 2N$.

The Hamiltonian (\ref{hamiltonian})
contains only marginal interactions -- this is the main,
though important, restriction to our study --
and includes several competing orders which are
encoded in the physical symmetry of the problem, the group H.
In Eq. (\ref{hamiltonian}),  these symmetries
are taken into account by the set of
symmetric matrices $D=\left\{ d^{\alpha} \right\}$
that commute with all elements of H:  $[d^{\alpha}, \mbox{H}]=0$.
On top of continuous symmetries,
the Hamiltonian (\ref{hamiltonian}) has discrete symmetries such as lattice
symmetries,  chiral
invariance,  charge conjugation and  parity.
In particular, those discrete symmetries are responsible
for the matrices
$d^\alpha$ to be real and symmetric.
In addition, model (\ref{hamiltonian})
should be stable and renormalizable under the RG approach and
we thus require the set of matrices $d^\alpha$
to be closed under anticommutation:
$\{ D,D\} \subset D$
\cite{leclair,wiese}.

In the following, we will generalize the problem
and make abstraction of the underlying fermions
to consider the G$_k$ WZNW model perturbed by current-current interaction
with Hamiltonian density:
\begin{equation}
\Cal H = \Cal H_0 + \Cal H_{\ind{int}} =
\frac{v_{\f}}{4\pi(k+\coxeter)} \;
\left({\cal J}_{\R}^A {\cal J}_{\R}^A + {\cal J}_{\L}^A {\cal J}_{\L}^A
\right)
+ \; g_{\alpha} \;{\J}_{\R}^A \; d^{\alpha}_{AB} \;{\J}_{\L}^B,
\label{hamiltonianref}
\end{equation}
where $A,B =1, \ldots, \mbox{dim} (G)$ and $\coxeter$ is the
dual Coxeter number of G (for instance $\coxeter=N$
for G = SU($N$) and $\coxeter=2N-2$ for
G = SO(2$N$)) \cite{difrancesco97}.
In this paper, G is a regular simple Lie group
and for application to weakly interacting fermionic models
(\ref{HamLatt}), one has
G =  SO(2N) with the KM level $k=1$
for model (\ref{hamiltonian}) at half-filling
and G =  SU($N$) and $k=1$ for
the spin degrees of freedom of model (\ref{hamspinsector})
away from half-filling.
The WZNW description (\ref{hamiltonianref}) has
to be supplemented by the defining G$_k$ KM current algebra:
\be
\J_{\L}^A(z)\J^B_{\L}(w)\sim \frac{k\delta^{AB}}{(z-w)^2} +
if^{ABC}\,\frac{\J_{\L}^C(w)}{z-w},
\label{opecurrent}
\ee
where a similar relation for right-moving currents
holds with the replacement $z,w\to\bar z,\bar w$.
In Eq. (\ref{opecurrent}), we choose a basis of the
Lie algebra $\frg$ of G for which the Killing form is the identity and
$f^{ABC}$ denote the structure constants of
$\frg$ which are normalized according to:
$f^{ABC} f^{ABD} = 2 \coxeter \delta^{CD}$.

In absence of interactions, i.e. when $g_\alpha = 0$, the
theory (\ref{hamiltonianref})  is
conformally invariant and its critical properties
are described by the G$_k$ WZNW CFT
with central charge $c(G, k) = k \; \mbox{dim} (G)/(k+\coxeter)$.
As far as global symmetries are concerned,
 the non-interacting model displays
a $\mbox{G}_{\L}\times\mbox{G}_{\R}$ symmetry,
which is generated by the charges
$Q^A_{\L(\R)}=\int dx\,\J^A_{\L(\R)}$.
One obvious effect of the interactions $g_\alpha\neq 0$,
on top of reducing the symmetry down to H,
is to couple the left and right sectors of the CFT:
the continuous part of the physical
symmetry group H of model (\ref{hamiltonianref}) is generated by a subset of
the \emph{diagonal} charges (i.e. that rotate
simultaneously the left and right sectors
of the theory) $Q^A_{\L}+Q^A_{\R}$, so that
 more rigorously
$\dr H
=\left(\mbox{H}_{\L}\times\mbox{H}_{\R}
\right)_{\ind{diag}}$.
For generic  $g_\alpha \neq  0$,
we will investigate
situations where the interaction is marginally relevant so that
a spectral gap opens and the conformal symmetry is lost.
\footnote{
An exception corresponds to the case where the marginal current-current
of model (\ref{hamiltonianref}) is described in terms of currents
with different KM levels. Even though the perturbation is marginally relevant
the conformal invariance is restored at an
intermediate infrared fixed point. The so-called chirally stabilized
spin-liquids are examples of this class of models \cite{adj,phlecsl}.
}
One important point about the interacting part
of the Hamiltonian (\ref{hamiltonianref}) is that it contains a
special, G-symmetric ray, i.e. there exist
some couplings $g_\alpha^0$ such that $g_\alpha^0d^\alpha = \idy$
($\idy$ being the $\mbox{dim}(\mbox{G})\times\mbox{dim}(\mbox{G})$
identity matrix).
If this were not the case, it would mean that the
Hamiltonian could be broken into smaller, commuting pieces,
built on subalgebras of G, to which our approach should be applied separately.
Along the G-symmetric ray, the Hamiltonian (\ref{hamiltonianref})
simplifies as follows:
\begin{equation}
{\Cal H}_{\mbox{G}} =
\frac{v_{\f}}{4\pi(k+\coxeter)} \;
\left({\cal J}_{\R}^A {\cal J}_{\R}^A +  {\cal J}_{\L}^A {\cal J}_{\L}^A
\right)
+ \; g\; {\J}_{\R}^A {\J}_{\L}^A,
\label{GNmax}
\end{equation}
which is the Gross-Neveu (GN) \cite{gross} or Thirring
model \cite{dashen} built on the group G at level $k$.
On this special ray, the global symmetry of
model (\ref{GNmax}) is given more rigorously by
$\dr G
=\left(\mbox{G}_{\L}\times\mbox{G}_{\R}
\right)_{\ind{diag}}$, which is
generated by the diagonal charges $Q^A_\L+Q^A_\R$.
This last remark finally provides us with the correct
way to introduce rigorously the group G, namely the maximal
symmetry group of the interacting model (\ref{hamiltonianref}).

\section{Emergent dualities}
\label{sectdual}

In this section, we study duality transformations
on the general model (\ref{hamiltonianref}).
After a brief survey of basic facts about the famous
KW duality in the quantum Ising model, we start by introducing
duality symmetries that map the set of theories defined by
Eq. (\ref{hamiltonianref}) onto itself, from a rather
algebraic point of view.
This definition will enable us to obtain a classification
of dualities depending on the nature of the group G.
Then, we study the interplay of such dualities
with the phenomenon of DSE. For readers not
interested in the technical details, a brief summary
of our main results is presented at the end of this section.

 %%%%%%%%%%%%%%%%%%%%%%%%%%%%%%%%%%%%%%%%%%%%%
 %%%%%%%%%%%%%%%%%%%%%%%%%%%%%%%%%%%%%%%%%%%%%

 \subsection{Dualities and covariance of the Hamiltonian}
 \label{dualcov}

%%%%%%%%%%%%%%%%%%%%%%%%%%%%%%%%%%%%%%%%%%%%%

Dualities are precious tools that help our
understanding of strongly correlated systems. Given a set of
theories depending on some parameters $\{g_\alpha\}$, with
Hamiltonian $H_{g_\alpha}[\phi]$ and fluctuating fields $\{\phi(x)\}$,
a duality $\Omega$ is a symmetry operation that relates
different points in this set of theories,
$H_{g_\alpha}[\phi]=H_{\tilde g_\alpha}[\tilde\phi]$, where
parameters as well as fields are acted upon by
the duality: $\Omega(g_\alpha)=\tilde g_\alpha$
and $\Omega(\phi)=\tilde \phi$ with $\Omega^2 = \idy $.
This ensures that the whole physical content of the
theory $H_{\tilde g_\alpha}$ can be deduced from
that of the theory $H_{g_\alpha}$:  correlators are
invariant, $\big\langle\hdots \tilde\phi(x)\hdots\big\rangle_{\tilde g_\alpha}
= \vm{\hdots \phi(x)\hdots}_{g_\alpha} $, where
$\vm{\hdots}_{g_{\alpha}}$ is the quantum average
defined by Hamiltonian $H_{g_\alpha}$.

\subsubsection{Warming up: Kramers-Wannier duality}

The 1D Ising model in a transverse magnetic
field provides us with maybe the most famous
example of such a duality, the so-called KW
duality. It is worth recalling basic facts about this model,
since it reveals striking similarities with the more
general situation studied later. Its lattice Hamiltonian is:
\be
H=\sum_i \left[g_1^\low\sigma^z_i\sigma_{i+1}^z + g_2^\low\sigma^x_i \right] ,
\ee
where $\sigma^a_i$ are Pauli matrices on site $i$. Defining the operators
$\Omega(\sigma_i^ z)=\mu^z_i=\prod_{j<i}\sigma^x_j$
and $\Omega(\sigma^x_i)=\mu_i^x=\sigma^z_i\sigma_{i+1}^z$ allows
a dual description of this model, $H=\sum_i \left[g_2^\low\mu^z_i\mu_{i+1}^z
+ g_1^\low\mu^x_i \right]$. The KW duality
$\Omega$ simply exchanges the couplings, $\Omega(g_{1,2}^\low)=g_{2,1}^\low$.
At the self-dual point $g_1^\low=g_2^\low$, the
theory is critical and described
by a $c=\frac{1}{2}$ CFT which is the free
Majorana fermion theory \cite{difrancesco97,bookboso}. Moving away from
the self-dual point amounts to adding a mass term
to the fermions, yielding the following Hamiltonian density
in the continuum limit:
\be
\Cal H = -\frac{iv}{2}\big(\xi_{\R}\p_x\xi_{\R}-
\xi_{\L}\p_x\xi_{\L} \big) + im\xi_{\R}\xi_{\L},
\label{ising}
\ee
with $m\propto g_1^\low-g_2^\low$. Then the KW duality exchanges the two
massive phases of the Ising model: when $m>0$, it is in
its ordered phase, $\vm{\sigma^z}\neq 0$, $\vm{\mu^z}= 0$,
while the disordered phase is characterized
by $\vm{\mu^z}\neq 0$ and $\vm{\sigma^z}= 0$. On top of
Majorana fermions, there are two other fluctuating
fields in the theory, $\sigma(x)$ and $\mu(x)$, that
are the continuum limits of the Ising spin $\sigma^z$,
and of the disorder operator $\mu^z$, respectively.
The KW duality has the following possible representation
on the continuous fields:
$\xi_{\L}\to-\xi_{\L}$, $\xi_{\R}\to\xi_{\R}$,
$\sigma\oto\mu$ which is indeed a symmetry
of model (\ref{ising}) with $m \rightarrow - m$.
Note also that if the original $\Z_2$ symmetry is
broken (e.g.~by adding a magnetic field along $\sigma^z$),
then the KW duality no longer holds. More precisely,
if one then requires the model to be globally invariant
under KW (up to a redefinition of the couplings), one
is forced to include a new term in the Hamiltonian,
namely a term proportional to $\mu^z$ which
is \emph{not local} with respect to the original
building blocks $\sigma^a_i$. In other word,
preserving the possibility of KW duality requires
to enlarge drastically the class of
perturbations. This is a general fact: as we will see,
for the generic model (\ref{hamiltonianref}), the set of
allowed dualities strongly depends on the symmetry group
of the model. Moreover, a quantum critical behavior
is also likely to emerge at the self-dual points.
In this respect, the KW duality approach has been fruitfully exploited
in the past to obtain the massive phases and quantum phase
transitions of two-leg spin ladders for weak
interchain interactions \cite{shelton}.
%%%%%%%%%%%%%%%%%%%%%%%%%%%%%%%%%%%%%%%%%%%%%

\subsubsection{Characterization and classification of dualities}

Motivated by the example of the Ising model, we now investigate
whether the theories defined by (\ref{hamiltonianref})
admit non-trivial transformations that relates
different points in this set of theories. Since the models
that are considered are built out of the current fields,
it is natural to define such a transformation
-- call it $\Omega$ -- by its action on the
currents, $\J\too \tJ= \Omega(\J)$. In close parallel
to the Ising model, we demand that the model
defined in terms of the new variables $\tJ$, corresponds
to the original one up to a redefinition of the
couplings, $g\too\widetilde{g}$:
\be
\Cal H (\tilde g,\tJ) = \Cal H(g, \J).
\label{defdual}
\ee
We will see shortly that such transformations can
exist, and, moreover, under very general
assumptions, are involutive; hence it is
sensible to coin them dualities.

In this paper, we will restrict ourselves to the
simplest case of transformations that act
linearly on the currents.
Imposing that the KM algebra (\ref{opecurrent})
be invariant prevents left-right mixing of
the currents, thus resulting in:
$\tJ_{\L(\R)}^A = (\omega_{\L(\R)})^{AB}\J^B_{\L(\R)}$,
$\omega_{\L(\R)}$ being $\mbox{dim(G)}\times\mbox{dim(G)}$
matrices representing the action of $\Omega$ on the currents.
From the invariance of the OPE (\ref{opecurrent}),
$\omega_{\L(\R)}$ are necessary orthogonal matrices.
By writing $\omega_{\L}=\omega \omega_{\R}$,
with $\omega=\omega_{\L}\omega_{\R}^{-1}$, we recognize
that the transformation can be decomposed
into the product of a pure \emph{diagonal}
G-rotation $\omega_{\R}$ (i.e.~that acts
simultaneously on the right- and left- chiral sectors)
and of a transformation acting only in the
left sector, $\omega$. It can be shown (see
Appendix \ref{app-dual-Cinv} for details) that
$\omega$ \emph{and} the diagonal rotation
$\omega_{\R}$ have to leave the Hamiltonian
globally invariant (i.e., they \emph{separately}
fulfill Eq. (\ref{defdual})). Therefore, the problem of
finding all possible dualities factors out
in two independent sub-problems.
The diagonal rotation $\omega_{\R}$ corresponds to a
mere change of basis in the representation
of the Hamiltonian. If the low-energy theory that one
considers is the continuum limit of some lattice model,
this diagonal rotation is the continuum
representation of a \emph{local}, unitary transformation
of the lattice operators. Here, we will not be
interested in these diagonal transformations.
We mention that
such transformations have been studied on the lattice
in the context of the
two-leg Hubbard ladder \cite{momoi03} and in the two-leg spin
ladder with ring-exchange interactions \cite{momoi4spin,totsuka}. One can
convince oneself that they appear whenever there is
some accidental degeneracy in the decomposition of the
Lie algebra of G in irreducible representations of H.
\footnote{An archetypical example of this is the spin-orbital model,
in which spin and orbital degrees of freedom are treated
on an equal footing, both transforming under SU(2) rotations,
with H=SU(2)$_{\mbox{orb}}\times$SU(2)$_{\mbox{spin}}$ $\subset$ G=SU(4).
Then, there is a SU(4) automorphism exchanging spin and
orbital degrees of freedom. We will come back
to this example in section V.}

In the following, we will investigate
the more interesting transformations that affect
only one chirality sector, setting therefore
$\omega_{\R}=\idy$. These transformations will be
called dualities, and generically denoted by $\Omega$.
The matrix $\omega$ is the representation of $\Omega$
on the KM current fields;  $\Omega$ twists
the current algebra:
\bea
\J_{\L}^A &\stackrel{\Omega}{\too}& \tJ_{\L}^A
=\omega^{AB}\J_{\L}^B,\nonumber\\
\J_{\R}^A &\stackrel{\Omega}{\too} &\tJ_{\R}^A
=\J_{\R}^A.
\label{twistdual}
\eea
Contrarily to the diagonal rotations, these transformations
that affect differently the left- and right-sector of
the theory cannot be viewed as the continuum limit of
a local transformation of the underlying lattice fermions.
In this sense, they are non-local, like the
KW duality in the Ising model which involves
non-local transformation of lattice operators.
Moreover, in general and contrarily to the KW
duality, there is no known, simple lattice
realization of the dualities in terms of fractional variables
(akin to the Ising spin and disorder operators
in the case of KW duality, that are fractional with
respect to the fermion)
these transformations are thus in
general \emph{emergent} in the continuum limit
\footnote{Note that our study does by no means
exclude the possibility of exact, complicated, necessarily
non-local lattice realizations of dualities in some
specific examples.}.

We now proceed to characterize the set $\du$ of possible
dualities $\Omega$. It turns out that the structure of
$\du$ is remarkably simply
extractible from few basis data: the physical symmetry
group H, and the maximal symmetry group G of the interaction.

First, one observes that $\omega$ should preserve
the KM current algebra (\ref{opecurrent}); it results that
$\Omega$ is an automorphism of $\frg$.
The covariance of the interaction term imposes further
constraints: there should exist
couplings $\tilde g_\alpha=\Omega(g_\alpha)$, such that
\be
g_\alpha \,d^\alpha=\tilde g_\alpha \, d^\alpha\omega.
\label{faldual}
\ee
Multiplying this relation by constant symmetric matrices
and taking a trace, we see that the transformed
couplings $\tilde g_\alpha$ are linearly related to
initial couplings $g_\alpha$. Moreover, condition (\ref{faldual})
is strong enough to completely characterize allowed dualities:
it is possible to show (see Appendix \ref{app-dual-Cinv}) that
the set of allowed dualities is given by:
\be
\du=\Cal C(\mbox H)\big|_{\ind{inv}} ,
\label{dualset}
\ee
i.e., the matrix $\omega$ should be the
representation (on the adjoint representation of G) of
those elements of the center of H
\footnote{We recall that the center of a group H is defined from
the elements which commute with all elements of H.},
that are
involutive, $\omega^2=\idy$.
This characterization of $\du$ has direct, simple
consequences. First, it shows that there is a
\emph{finite} number of dualities. $\du$ is indeed
isomorphic to the finite group $(\Z_2)^n$, $n$
being some positive integer, and has $2^n$ elements.
Furthermore, to each duality, there corresponds
a symmetry operation $\Cal S$ that belongs to the
physical symmetry group H, which has the following
properties: $\Cal S^2=\idy$ ($\Cal S$ is
a $\Z_2$ symmetry), and $\Cal S$ commutes
with all other symmetry operations of H. Explicitly, a
representation of $\Cal S$ can be given as:
$\Cal S=\omega\times\omega$, where the two operands of
the tensor product act on the left- and right- sector
of the theory, respectively. A duality therefore corresponds
to the ``square root'' of some exact involutive
symmetry of the lattice Hamiltonian that commutes
with all other symmetries.

Let us denote by $\Omega_a$ the different
dualities, $\du=\ens{\Omega_a}{}$. Using the
properties $\com{\omega_a}{\omega_b}=0$
and $\omega_a^2=\idy$, we deduce that there exists
a basis for the currents that simultaneously
diagonalizes the action of dualities on the
currents, the eigenvalues being $\pm 1$. In the following we
will work in such a basis:
\bea
\Omega_a(\J^A_{\L}) &=& \epsilon_a^A\, \J_{\L}^A,
\qquad  \epsilon_a^A=\pm 1.
\label{repdual}
\eea
This yields a more physical interpretation of dualities.
Introducing Noether charge and current densities associated
to the G-invariance, $J^A_{0}=\J^A_{\R}+\J^A_{\L}$
and $J^A_1=\J^A_{\R}-\J^A_{\L}$ (the subscripts correspond
to space-time indices, $x_0\equiv t$ and $x_1\equiv x$,
and in a G-invariant theory $\p_t J^A_0+\p_x J^A_1=0$), they
transform as:
\bea
\Omega_a(J^A_0)=J^A_0 ,\quad\Omega_a(J^A_1)=J^A_1,
\qquad \epsilon^A_a=+1,\nonumber\\
\Omega_a(J^A_0)=J^A_1, \quad\Omega_a(J^A_1)=J^A_0 ,
\qquad \epsilon^A_a=-1.
\label{noether}
\eea
In a good basis, a duality just amounts to
the exchange of the role played by Noether charges and
currents associated to G-invariance. Thus, it is
tempting to view dualities as generalizations of
the U(1) duality that exchanges the role played by electric and magnetic charges
(this is known as "T-duality" in the context of string theory \cite{giveon94}), to a non-Abelian theory,
with the further constraint that they have to be
compatible with the original H-invariance of the microscopic theory.
This will have important consequences for the labeling of
states by quantum numbers. Indeed, if the eigenvalues of the
charge operator $Q^A_0=\int dx\, J^A_0(x)$ happen to
be (approximate) good quantum numbers in a phase $\M$,
and if $\epsilon^A_a=-1$, it follows that the (approximate)
good quantum numbers in the dual phase $\Omega_a(\M)$ will
be \emph{current} quantum numbers: quasiparticles
carry generalized currents.

We now make use of a mathematical result on
Lie algebras that allows for a complete classification
of dualities. Given $\Omega_a\in\du$, using
Eq. (\ref{repdual}), it is possible to decompose
the Lie algebra $\frg$ of G into two orthogonal
subspaces: $\frg=\frg_\parallel^a\oplus\frg_\perp^a$,
$\frg_\parallel^a$ ($\frg_\perp^a$ respectively) being
generated by those elements of $\frg$ with
$\epsilon^A_a=+1$ ($\epsilon^A_a=-1$ respectively).
Mathematically, this decomposition can be associated
to any involutive automorphism ($\omega$ in our case),
which is called a
$\Z_2$-grading of $\frg$ \cite{fuchs}.
Each $\Z_2$ grading $\Omega$ is
characterized by an invariant subspace $\frg_\parallel$,
which can be shown to have the structure of a semi-simple
 Lie algebra, and is defined by
$\Omega(X)=X,$ $\forall X\in\frg_\parallel$. One has the
orthogonal decomposition (with respect to the
Killing form)  $\frg=\frg_\parallel\oplus\fr{p}$,
with $\Omega(X)=-X,$ $\forall X\in\fr{p}$, and the Lie
structure of $\frg$ breaks up as follows under
the action of $\Omega$:
\be
\com{\frg_\parallel}{\frg_\parallel}=\frg_\parallel,\quad
\com{\frg_\parallel}{\fr{p}}=\fr{p},\quad
\com{\fr{p}}{\fr{p}}=\frg_\parallel.
\ee
Now, there exists a complete classification
of $\Z_2$-gradings for simple
Lie algebras \cite{helgason}.
This means that once G is given, irrespective
of the physical symmetry group H, one knows what are
the different possible types of dualities: they are
of the form $\omega=U \bar \omega U^{-1}$, where
$U$ belongs to G, and where $\bar \omega$ is one
representative of the possible, classified,
$\Z_2$-gradings of $\frg$.
For regular simple
Lie groups G, the $\Z_2$-gradings are presented
in Table \ref{tabZ2} \cite{helgason}.

\begin{table}
\begin{center}
\begin{tabular}{|c|c|c|}
\hline
Type & G & $\frg_\parallel$\\
\hline
A\I & SU($N$) & $\mathfrak{so}$($N$) \\
\hline
A\II & SU($2N$) &  $\mathfrak{sp}$($2N$)\\
\hline
A\III & SU($p+q$) & $\mathfrak{s}(\mathfrak{u}(p)\oplus\mathfrak{u}(q))$
\\
\hline
BD\I & SO($p+q$) & $\mathfrak{so}(p)\oplus\mathfrak{so}(q)$\\
\hline
D\III & SO($2N$) & $\mathfrak{u}(N)$\\
\hline
C\I & Sp($2N$) & $\mathfrak{u}(N)$\\
\hline
C\II & Sp($2p+2q$) & $\mathfrak{sp}(2p)\oplus\mathfrak{sp}(2q)$\\
\hline
\end{tabular}
\end{center}
\caption{Exhaustive list of $\Z_2$-gradings for simple Lie algebras.
For the $\Z_2$-grading of the A\III type, we use
a somewhat non-standard notation to designate
the algebra of block diagonal matrices, with blocks
$A$ and $B$ on the diagonal, where
$A$ (respectively $B$) is a $p\times p$ (respectively $q\times q$)
Hermitian matrix with the constraint
Tr$(A+B)=0$. In this case, $\frg_\parallel$ is isomorphic to
$\mathfrak{su}(p)\oplus\mathfrak{su}(q)\oplus \mathfrak{u}(1)$.
}
\label{tabZ2}
\end{table}

In section IV,
we will investigate in detail the case G=SU($N$)
for applications to fermionic models (\ref{hamiltonian}).
In this picture, the role of the physical symmetry group H is to
select what are the realized dualities amongst the set of
possible ones. That the set $\du$ of dualities is not reduced to
the trivial identity element will be clear
later on the example presented in section \ref{secexample}.
%%%%%%%%%%%%%%%%%%%%%%%%%%%%%%%%%%%%%%%%%%%%%

\subsubsection{Transformation of fields}

So far, one has fully characterized the set of
dualities $\du$, that depends only on the maximal
symmetry group G and the physical symmetry group H.
This was done by investigating the action of
dualities on particular fields, the left-moving
currents of the theory (the action on the right
currents being trivial). To investigate the effect
of dualities on observables, we need more generally
their action on any local field of the
G$_k$ WZNW model (the unperturbed theory); this action turns out
to be particularly simple.

The whole field content of the WZNW model can be recovered
by considering the KM primary
operators $\Phi_{\lambda_{\L},\lambda_{\R}}$ \cite{difrancesco97}.
Since the WZNW model is invariant under independent
global \emph{chiral} G-rotations in the right and
the left sectors,
each primary field is labeled by two G-highest weights
$\lambda_{\L}$ and $\lambda_{\R}$ that dictate how
the primary field transforms under the group
G$_{\L}\times$G$_{\R}$. Each primary field is thus
a tensor $\left(\Phi_{\lambda_{\L},\lambda_{\R}}\right)_{a_{\L},a_{\R}}$,
$a_{\L(\R)}=1,...,\mbox{dim}(\lambda_{\L(\R)})$.
The precise set of representations $\lambda_{\L(\R)}$
that occurs in the WZNW model defines the field content
that depends on the level $k$ \cite{difrancesco97}.

Before we give the action of dualities on these primary
fields, we need a few basic facts about
the group of automorphisms Aut(G). It contains the subgroup
of \emph{inner} automorphisms Aut$_0$(G), that can be
identified with conjugation by elements of G, i.e.
those transformations $A_U$ that act on G as:
$g\too A_U(g)=UgU^{-1}$, with $U\in$ G. Thus, inner
automorphisms can be viewed as the group of changes
of basis in the representations of G
\footnote{Strictly speaking, Aut$_0$(G) contains
more than conjugation by elements of G, since it is
the connected component of the identity in Aut(G). This
subtlety will play no role in our study.}.
In fact, inner automorphisms do not exhaust Aut(G).
Indeed, Aut(G)/Aut$_0$(G) is in general a non-trivial discrete group.
Whenever it is not trivial, we will denote its non-trivial
elements generically by $\tau$. This leads us to
introduce \emph{outer} automorphisms, that can be
written as $A\circ\tau$ with $A\in$Aut$_0$(G), and
to write Aut(G)=$\big\{$Aut$_0$(G),Aut$_0$(G)$\circ\tau\big\}$.
Contrarily to inner automorphisms, that leave globally
invariant any G-representation, outer automorphisms
generically \emph{exchange} representations,
i.e.~$\tau(\lambda)\neq\lambda$. Outer automorphisms are in
one-to-one correspondence (up to conjugation) with
the symmetries of the Dynkin diagram
associated to G \cite{fuchs}.

For regular simple Lie algebras, involutive outer
automorphisms do exist and will be relevant to our
study (they will give rise to ``outer'' dualities). An
exhaustive list of outer automorphisms is: for G=SU($N$), the
non-trivial element $\tau$ corresponds to \emph{charge conjugation},
that maps a representation on its charge conjugated
partner  -- the corresponding outer $\Z_2$-gradings
are of type A\I and A\II (see Table I) -- ; for G=SO($2N$),  the non-trivial
element $\tau$ corresponds to \emph{``SO(2N) parity''}, that
exchanges the two spinorial representations
--  the corresponding
outer $\Z_2$-gradings is of type BD\I when $p$ and $q$ are
odd (see Table I) --.

Let us now describe the action of dualities on fields.
Since a duality $\Omega\in\du$ affects only the
left sector, its action on the primary
field $\Phi_{\lambda_{\L},\lambda_{\R}}$ is readily found:
it is simply given by the action of the automorphism
$\omega$ on the left representation $\lambda_{\L}$.
If $\Omega$ belongs to the inner class, it just amounts
to a change of basis in each representation
$\lambda_{\L}$. Denoting by $U_{\lambda_{\L}}$ the matrix
performing the change of basis, the primary operator becomes:
\be
\Phi_{\lambda_{\L},\lambda_{\R}} \xrightarrow{\;\Omega\;}
\tilde \Phi_{\lambda_{\L},\lambda_{\R}}=
U_{\lambda_{\L}}\cdot\Phi_{\lambda_{\L},\lambda_{\R}}
\esp \Omega\mbox{~~inner}.
\label{primaryInn}
\ee
In this expression,
a matrix product is implied; in components it
reads as follows: $\big(\tilde\Phi_{\lambda_{\L},
\lambda_{\R}}\big)_{a_{\L},a_{\R}}=
\left(U_{\lambda_{\L}}\right)_{a_{\L} b_{\L}}
\left(  \Phi_{\lambda_{\L},\lambda_{\R}}\right)_{b_{\L},a_{\R}}$.
On the other hand, outer dualities map in general
the representation $\lambda_{\L}$ on another
representation $\lambda_{\L}^*=\tau(\lambda_{\L})$, and
also perform a change of basis in $\lambda_{\L}^*$. Thus,
the primary operator transforms under duality according to:
\be
\Phi_{\lambda_{\L},\lambda_{\R}} \xrightarrow{\;\Omega\;}
\tilde \Phi_{\lambda_{\L},\lambda_{\R}}=
U_{\lambda_{\L}^*}\cdot\Phi_{\lambda_{\L}^*,\lambda_{\R}}
\esp \Omega\mbox{~~outer}.
\label{primaryOut}
\ee

Let us now briefly restate the main results of the preceding sections:
\begin{itemize}
\item To each family of models (\ref{hamiltonianref})
defined by (i)
an unperturbed critical model, namely a WZNW model
built on the symmetry group G, at level $k$ (for the fermionic
models (\ref{hamiltonian}), the level is fixed to $k=1$)
and
(ii) its bare symmetry group H, there is a finite
set of exact dualities $\du$.

\item To each model  $\M$ belonging to this family,
defined by its bare couplings $g_\alpha$, one can
associate a finite number of models $\M_\Omega$ defined
by couplings $\Omega(g_\alpha)$ where $\Omega\in\du$
is an acceptable duality. These dualities are \emph{exact}
in the continuum limit. The physical properties of
$\M_\Omega$ can in principle entirely be deduced
from those of $\M$. For example, suppose the ground state
of model $\M$ is specified by some order parameter $\Cal O$,
that can be expressed in terms of a primary operator:
$\Cal O=\tr\left(M\Phi_{\lambda_{\L},\lambda_{\R}} \right)$, where
$M$ is some matrix encoding the ``texture'' of the
phase of model $\M$. Then, the ground state for model
$\M_\Omega$ is immediately given, with the help
of Eqs. (\ref{primaryInn},\ref{primaryOut}),
by $\Cal O_\Omega=\tr\left(M\tilde \Phi_{\lambda_{\L},\lambda_{\R}} \right)$.
\end{itemize}

It follows that it would be sufficient to
understand the physics of one ``fundamental chamber"
-- a subspace of the space
of the couplings $g_\alpha$ -- from which physical
properties for the whole space of couplings could
be deduced by acting with dualities. Nevertheless,
understanding the physics of the ``fundamental chamber"
is still of course a complicate task. In particular, there is no
reason for this ``fundamental chamber" to display a single
type of physical behavior -- it may contain many
different phases. In section IV, we will give
a physical picture of part of
the ``fundamental chamber" in the SU($N$) case to
exhaust its physical properties.

%%%%%%%%%%%%%%%%%%%%%%%%%%%%%%%%%%%%%%%%%%%%%
%%%%%%%%%%%%%%%%%%%%%%%%%%%%%%%%%%%%%%%%%%%%%

\subsection{Dualities and DSE}
\label{dualDSE}

In this section, we present an alternative characterization
of the dualities found in
the previous section:
they are in one-to-one correspondence with the different ways
the bare physical symmetry group H of the model can be
promoted to a larger dynamical symmetry group G. This notion of
DSE phenomenon, which can be made explicit
by means of the RG approach \cite{konik02}, is recalled for
completeness and applied
to model (\ref{hamiltonianref}). Then, important consequences
for the global phase diagram
are drawn, that allow for a general picture thereof.

%%%%%%%%%%%%%%%%%%%%%%%%%%%%%%%%%%%%%%%%%%%%%

\subsubsection{DSE phenomenon}

Perturbing a 1D critical model
with a continuous symmetry offers in
general a whole family of possibilities, that one can
label by the symmetry breaking pattern and by
the type of perturbing operator; in our case,
once chosen the general class of perturbations of
the current-current type (\ref{hamiltonianref}), the model is specified
by the pattern G$\downto$ H. Generically, a
marginal relevant perturbation results in the opening of a
spectral gap in \emph{every} sector of the theory.
However, in one dimension,
a simplification can occur that might allow for a kind
of universal description of these fully gapped phases.
As we have seen in section II,  the family of models (\ref{hamiltonianref})
contains a special, isotropic ray (with
couplings proportional to $g_\alpha^0$), where model (\ref{GNmax})
displays an extended global G symmetry. This GN model,
that we will call $\M_0$, has the following Hamiltonian density:
\be
\Cal H_G = \Cal H_0 + g\,\J^A_{\R}\J^A_{\L} .
\label{hamM0}
\ee
For $g >0$, this model is
a massive integrable field theory and
its spectrum is known for
$k=1$ \cite{andrei79,zamolodchikov79,karowski81} and in the
general $k$  case \cite{ahn90,hollowood,babichenko}.
The coupling $g$ flows
to strong coupling under RG, and a mass scale
is dynamically generated: $m\propto e^{- {\rm cst}/g}$.

It turns out that physical properties of the
generic model (\ref{hamiltonianref}) can be
weakly sensitive to anisotropy, i.e. to departure
from the isotropic ray, in a sense we are going to discuss now.
To restate it differently, the effective theory
describing the bare anisotropic model at
low energy, displays approximately a \emph{larger}
symmetry G, and the physics is that of the
maximally symmetric model $\M_0$. This phenomenon has
been coined DSE.
An indication for
DSE can be gained from considering the
RG beta-function for generic models of the
form (\ref{hamiltonianref}). As shown in the detailed
study of Refs. \onlinecite{konik02} and \onlinecite{lin}, the beta-function
generically indicates that the more symmetric
ray is attractive under RG (writing $g_\alpha
= g g^0_\alpha+\delta g_\alpha$ the
quantity $\delta g_\alpha/g$ flows to
zero at one loop
$\lim_{\ell\to 0}(\delta g_\alpha/g)= 0$,
with $\ell=\ln(\Lambda a_0/v_\f)$ the RG time, $\Lambda$
the running cut-off and
$a_0$ the lattice spacing).
The model thus flows to more and
more symmetric theories at low energies. The RG flow
defines a scale $\Lambda_\DSE$, akin to Josephson length,
where anisotropy becomes negligible:
$(\delta g_\alpha/g)(\ln(\Lambda_\DSE a_0/v_\f))\ll 1$.
Now focusing on the low-energy spectrum (precisely speaking at energies
$E<E_{\ind{low}}\ll\Lambda_\DSE$), we can have a non-trivial
regime described by
field theory only if one considers the triple limit
\be
\mbox{(strong DSE)\hspace*{2cm}}
m\ll E_{\ind{low}}\ll\Lambda_\DSE \ll v_\f a_0^{-1} ,
\label{universal}
\ee
where $m\propto e^{-\ind{cst}/g}$ is the typical gap
scale of the system (in this
asymptotic DSE regime a single coupling $g$ is enough
to describe the flow).
Regime (\ref{universal})  is \emph{universal}:
in that limit the lattice model (\ref{HamLatt}) and model
(\ref{hamM0}) share the same
\emph{quantitative} physical low energy ($E<E_{\ind{low}}$) properties (this
quantitative DSE phenomenon could be checked explicitly
in some cases where the anisotropic model
itself is integrable, allowing for a complete
determination of all
physical quantities \cite{konik02}).
It is not very surprising that when dealing with
gapped systems, the notion of  universality requires far
more restrictive conditions -- as far as the
regime of bare parameters of the underlying lattice model
is concerned -- than for critical systems
(for which a Josephson length $\Lambda_J$ can also be
defined and that only require $E_{\ind{low}} \ll \Lambda_J \ll
v_\f a_0^{-1}$). Note that (\ref{universal}) requires
in particular $m\ll\Lambda_\DSE$, which implies that
the running coupling constants are small at the DSE scale:
$g_\alpha(\Lambda_\DSE)\ll 1$. This shows that strong DSE is
essentially a weak-coupling phenomenon \cite{konik02,lin}, and
this justifies the one-loop approximation.

It is also possible to define DSE in a
weaker sense, in which one does not require
that the physical quantities be quantitatively
the same as those of the isotropic point.
The meaning of DSE in the weak sense, can maybe be best
grasped in light of adiabatic continuity:
it simply reflects the fact that the ground state
at the isotropic point can be adiabatically
connected to the ground state of the (weakly)
anisotropic models; moreover, this adiabatic
continuity also holds for the quasiparticles
that can still be labelled by G quantum numbers:
each of the G-multiplets $\lambda$ of quasiparticles
of the isotropic GN model splits under the (small) residual
anisotropy breaking, $m_\lambda\to(m_\lambda^1,m_\lambda^2,...)$,
with a splitting
$\delta m_\lambda
= \mbox{max}(|m_\lambda^i-m_\lambda^j|)\ll m_\lambda$.
This gives the hint that if one does not insist
on the physical quantities to be asymptotically \emph{identical}
to those on the isotropic ray, then one can allow
for finite, small deviations from the limit
$m/\Lambda_\DSE\to 0$.
This has been shown in the
particular case of the SU(4) Hubbard chain at half-filling \cite{assaraf04},
where continuity could be checked numerically at finite anisotropy
in a non-integrable model.
Thus, DSE in the weak sense simply means that the
group G is an approximate symmetry of the low-energy
sector of the theory, that allows for a description
of the low-energy sector -- very much in the way
approximate symmetries allow for a classification
of baryons and mesons in particle physics.
As shown in the following section, the splitting of
the G-multiplet under the residual anisotropy can be
controlled in a perturbative expansion, which means
in particular that the quasiparticles of the isotropic
point (of model $\M_0$) survive anisotropy
\footnote{Of course, this cannot be the case
irrespective of the perturbation type: our statement
holds provided the anisotropy is due to current-current terms.},
and establishes the aforementioned adiabatic
continuity on a firm ground.
This weak-DSE regime,
corresponding to:
\be
\mbox{(weak DSE)\hspace*{2cm}} m\lesssim \Lambda_\DSE
\label{weakDSE}
\ee
is characterized by adiabatic continuity of the low-energy
spectrum (in particular, it is a smooth crossover and the
symmetry of the ground state cannot change).
Of course, symmetry enlargement cannot occur
everywhere in the phase diagram: it has been
established on the level of a perturbative
expansion around the isotropic ray, and if the
bare anisotropy is too strong, one expects it to break down.
In particular, when $\Lambda_\DSE/m$ decreases and $\delta g/g$
becomes of order one, level crossings start to occur in the
lowest energy sector of the theory
and the labeling in terms of G quantum
numbers losses its meaning.
In the following section, we will actually see that the existence
of dualities, that where introduced in section \ref{dualcov}
purely on group theoretical considerations,
implies that is \emph{has} to break down, and in a rather
elegant way, since it will
allow to grasp at a glance the (very rough) structure of the
phase diagram.

%%%%%%%%%%%%%%%%%%%%%%%%%%%%%%%%%%%%%%%%%%%%%%%
%%%%%%%%%%%%%%%%
\subsubsection{Interplay with dualities}

We are now in position to make the connection
with the dualities introduced in the
previous section. It is possible to sketch
the phenomenon of DSE as follows:
\be
\mbox{G}_{\L}\times\mbox{G}_{\R} \xrightarrow{\;\;\ind{interaction}\;\;}
\dr H=\left(\mbox{H}_{\L}\times\mbox{H}_{\R}\right)_{\ind{diag}}
\xrightarrow{\ind{\hspace{0.5cm}DSE\hspace{0.35cm}}} \dr G^{I\!R},
\label{schemaresto}
\ee
where the global symmetry group is indicated
(from left to right, respectively) before
perturbation, in the bare H-symmetric
model (\ref{hamiltonianref}), and in the effective
low-energy theory in the infrared (IR) regime after DSE
has occurred for instance in the weaker sense
with the symmetry group $G^{I\!R}$.
One can then ask the following
question: is diagram (\ref{schemaresto}) unique?

It turns out that the answer is positive up to
twists in the current algebra of the
form (\ref{twistdual}), i.e., up to dualities.
Indeed,
the final point of diagram (\ref{schemaresto}) is in general of the form
$\mbox{G}^{I\!R}=\big(\tilde{\mbox{G}}_{\L}
\times\mbox{G}_{\R}\big)_{\ind{diag}}$
with generators given by
$\int dx\,\big( \tJ^A_{\L}+\J^A_{\R}\big)$,
where $\tJ^A_{\L} = \omega^{AB}\J^B_{\L}$
and $\omega\in\du$ is a duality
 (see Appendix \ref{app-dual-dse} for a proof).
Dualities can thus be viewed as the different
possible DSEs compatible with a given bare
symmetry group H. To each duality $\Omega\in\du$,
there corresponds a generalized
isotropic, $\tilde{\mbox{G}}$-symmetric ray, described
by the following Hamiltonian density:
\be
\Cal H_\Omega = \Cal H_0 +g\,\J^A_{\R}\,\omega^{AB}\,\J_{\L}^B ,
\label{hamOmega}
\ee
a model that we will denote
$\M_\Omega$. Being related by a duality
to the fundamental isotropic model (\ref{hamM0}), it is
also an integrable field theory.

\subsubsection{Symmetry enlarged phases}

Since dualities are exact isometries of the beta-function,
model (\ref{hamOmega}) is also attractive
in the RG sense, so that DSE also occurs in the
vicinity of the (generalized) isotropic ray. Now, the fact
that DSE breaks down if the bare anisotropy is too
large acquires a new meaning: it has to be so, since
the different DSE schemes corresponding to
different dualities are incompatible; in other words,
the different symmetry enlarged phases
$\M_\Omega$ compete. We will turn shortly to
the study of this competition.

What about the robustness of DSE, i.e. what
happens when one moves away,  in the space of
bare couplings,  from the isotropic ray
$g^\Omega_\alpha=\Omega(g_\alpha^0)$ (with
$g^\Omega_\alpha d^\alpha = \omega$, see
Eq. (\ref{faldual}))? One way to
answer this question is to consider the
one-loop RG flow, with initial conditions
$g_\alpha(a_0^{-1}) = g(a_0^{-1})
(g_\alpha^\Omega + \delta_\alpha(a_0^{-1}))$,
$\delta_\alpha\ll 1$. The latter has to be
cut-off before the couplings become of order
one, i.e. at a scale $\Lambda > m$, where
$m$ is  the mass scale. Departure from the
isotropic ray then results in a \emph{residual}
anisotropy, the couplings being
$g_\alpha(\Lambda) = g(\Lambda)
(g_\alpha^\Omega + \delta_\alpha(\Lambda))$. The study
of Ref. \onlinecite{konik02} ensures
that $\delta_\alpha(\Lambda)$ scales to 0
in the scaling limit $\Lambda a_0<<1$ when
approaching the isotropic ray, resulting in
corrections of order $\left(g(a_0^{-1})
\delta_\alpha(a_0^{-1})\right)^\beta$ in the
physical quantities, with $\beta>0$ an exponent
that depends on the groups G and H under consideration.

An alternative argument, that establishes
the adiabatic continuity of the low-energy physics
when departing from the isotropic rays, can be given
using on the integrability of
the isotropic model (\ref{hamOmega}).  At scale
$\Lambda$, the model can be described by
the following Hamiltonian density:
\be
\Cal H=\Cal H_\Omega +
g(\Lambda) \delta_\alpha(\Lambda)\, \J^A_{\R} d^\alpha_{AB} \J^B_{\L},
\label{hamAnisot}
\ee
where the last term is the residual anisotropy.
It is possible to investigate, by form factor methods,
the effect of this term which constitutes
a perturbation on top of an integrable model \cite{mussardo}:
if the perturbation is local with respect
to the quasiparticles, then
the quasiparticles survive the perturbation,
i.e. there is adiabatic continuity in the spectrum \cite{mussardo}.
Since the perturbation is built on currents,
that are the local generators of conserved charges,
one concludes to mutual locality between the
perturbation and the quasiparticles. It results
that the low-energy spectrum of
model (\ref{hamAnisot}) is described by the
quasiparticles of the integrable model (\ref{hamOmega}).

\subsubsection{Quantum phase transitions}
\label{secQPTgen}

Let us now consider two different symmetry
enlarged phases. The low-energy physics of each of
them is captured by generalized isotropic models
of the form (\ref{hamOmega}), to be
called $\M_{\Omega_1}$ and $\M_{\Omega_2}$. These two models
are exchanged by the duality $\Omega=\Omega_1\Omega_2$
(recall that dualities form a group).
In the space of bare couplings, amongst the
different trajectories connecting these two isotropic
rays, there is a special, maximally
symmetric one, that allows to study, in a minimal model,
the competition between the two
phases $\M_{\Omega_1}$ and $\M_{\Omega_2}$. Again,
we invoke DSE to argue that this maximally
symmetric trajectory is attractive under RG.

Recalling that the duality $\Omega$ corresponds
to a $\Z_2$ grading of
the Lie algebra of G ($\frg=\frg_\parallel\oplus\frg_\perp$),
there exists a basis $\{\J_\parallel^a,\J_\perp^b \}$
for the currents with the following properties:
\be
\Omega(\J_{\parallel \L(\R)}^a)= \J_{\parallel \L(\R)}^a \qquad
\Omega(\J_{\perp \L(\R)}^b)= \mp \J_{\perp \L(\R)}^b,
\ee
where $a=1,\ldots, \mbox{dim}(\frg_{\parallel})$ and
$b=1,\ldots, \mbox{dim}(\frg_{\perp})$.
Then, the maximally symmetric model interpolating
between the phases $\M_{\Omega_1}$
and $\M_{\Omega_2}$ has symmetry G$_\parallel$ and is given by
\footnote{In the case G$_\parallel$ is not a
simple Lie group, there are in general
several couplings $g_{\parallel,\alpha}$.
This does not affect our analysis.}:
\be
\Cal H_{1-2} = \Cal H_0 + g_\parallel\,
\J_{\parallel \R}^a\omega_1^{aa'}\J_{\parallel \L}^{a'}
+ g_\perp\, \J_{\perp \R}^b\omega_1^{bb'}\J_{\perp \L}^{b'}.
\label{minmodel}
\ee
When choosing $g_\parallel=g_\perp$, one recovers
model $\M_{\Omega_1}$, whereas $g_\parallel
=-g_\perp$ yields model $\M_{\Omega_2}$.
There is a special point where the competition between
the orders is at its climax: the self-dual point $\M^*_{1-2}$
corresponding to $g_\perp=0$, or, in term of the
original couplings, to $\Omega(g_\alpha)=g_\alpha$.

The model (\ref{minmodel}) can be further
simplified by acting with the duality $\Omega_1$:
this amounts to study the competition between the
phase $\M_{\idy} = \M_0$ and $\M_\Omega$, and
in (\ref{minmodel}) to replacement $\omega_1\to\idy$,
which will be assumed in the following.
Then, at the self-dual point, the Hamiltonian density
takes a simpler form:
\be
\Cal H^*_\Omega=\Cal H_0 + g\, \J_{\parallel \R}^a
\J_{\parallel \L}^a,
\label{selfdual}
\ee
i.e., the currents $\J_\perp$ do not appear in
the interaction term. This means that there is room
for criticality at the self-dual point,
with the decoupling of some degrees of freedom. The
currents $\J_{\parallel}^a$ generate an
affine algebra $\frg_\parallel$ at
level $k^*=\iota k$, where $\iota$, a positive integer, is the Dynkin
index for the embedding $\frg_\parallel\subset\frg$
\cite{difrancesco97}.
It is then possible to separate the ``parallel''
and ``perpendicular'' degrees of freedom at the level
of the non-interacting Hamiltonian:
$\Cal H_0=\Cal H_0(\frg,k)=\Cal H_0(\frg_\parallel,k^*)
+ \Cal H_{\Cal A}$, where the last piece is the
Hamiltonian of model $\Cal A$ describing the degrees
of freedom that decouple at the self-dual point.
Model $\Cal A$ (that can be trivial, i.e.,
with vanishing central charge), is the
coset model
\be
\Cal A = \frac{G_k}{G_{\parallel k^*}},
\label{defcosetA}
\ee
with central
charge $ c_{\Cal A}^{\vphantom{\dagger}}
= c(G,k)-c(G_\parallel,k^*)$,
where  we recall that $c(G,k)= k \; \ind{dim(G)}/(k+\coxeter)$
is the central charge of the G$_k$ WZNW model.
Model $\Cal A$ can be non-trivial, with non-integer
central charge. Concrete examples in the
context of fermionic models will be given
in sections \ref{secferm} and \ref{secexample}.
Of course, $c_{\Cal A}^{\vphantom{\dagger}}$
is only an upper bound to the central
charge of the ``perpendicular'' criticality at the self-dual point: one has
to check that no other relevant operator in
model $\Cal A$ is allowed by symmetries, that
would spoil criticality. One also has to check whether the parallel
sector itself can develop criticality.

\begin{figure}[ht]
\begin{center}
\vspace{0.5cm}
\includegraphics[angle=0,width=0.5\linewidth]{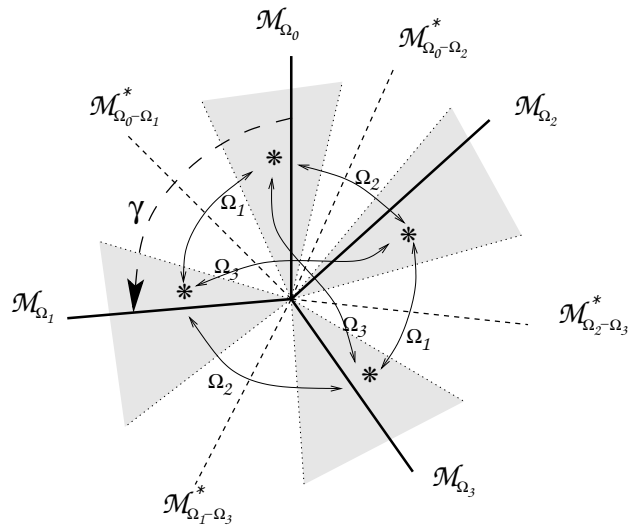}
\caption{Sketch of (a 2D-projection of) the phase
diagram for the example of a model
with four dualities:
$\Omega_0=\idy$, $\Omega_1$, $\Omega_2$,
and $\Omega_3=\Omega_1\Omega_2$. Bold lines
correspond to symmetry enlarged rays $\M_\Omega$,
dashed ones to self-dual models $\M^*_{\Omega-\Omega'}$
(for clarity some of the self-dual models
have been omitted). All lines meet at
the G$_k$ WZNW critical point (i.e. non-interacting
point). Gray areas indicate regions of DSE
and the dotted lines bounding it marks
a cross-over to a regime where DSE
looses its meaning (i.e. the splitting of
G-multiplets due to anisotropy becomes of
the order of the gap scale $m$). Four representative
points together with the dualities
exchanging them are also depicted.
The dashed curve $\gamma$ symbolizes
 the path in parameter space which
corresponds to the minimal  theory (\ref{minmodel}) interpolating between
phases $\M_\idy$ and $\M_{\Omega_1}$. See figure \ref{sketchcrossover}
for a qualitative picture of the low-energy spectrum along this path.}
\label{fig-phasegen}
\end{center}
\end{figure}

\begin{figure}[h]
\begin{center}
\vspace{0.5cm}
\includegraphics[angle=0,width=0.5\linewidth]{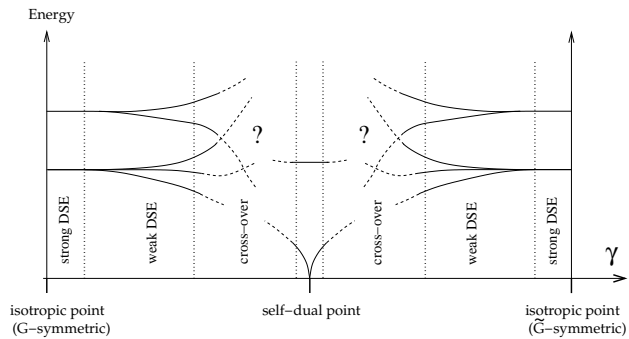}
\caption{Sketch of the low-energy spectrum of the minimal
model (\ref{minmodel}) interpolating between two symmetry
enlarged phases (see the path $\gamma$ in figure
\ref{fig-phasegen}).}
\label{sketchcrossover}
\end{center}
\end{figure}

We thus arrive to the following general
picture of the phase diagram for
model (\ref{hamiltonianref}) -- see figure \ref{fig-phasegen}:
\begin{itemize}
\item There is a finite set of generalized isotropic
rays where the bare model displays an \emph{exact}
enlarged $\tilde{\mbox{G}}$-invariance. Each
ray $\M_\Omega$ is labelled by a
duality $\Omega\in\du$, and the global
invariance group $\tilde{\mbox{G}}$ is obtained
from $\mbox{G}=\big( \mbox{G}_{\L}
\times \mbox{G}_{\R} \big)_{\ind{diag}}$ by
acting with $\Omega$ on the currents (see Eq. (\ref{twistdual})).
The Hamiltonian on those rays is
given by Eq. (\ref{hamOmega}).

\item Each of those isotropic rays is
attractive in the RG sense, and to it there
corresponds a pocket with \emph{finite}
extension in the space of bare couplings.
In this pocket, the low-energy sector of the
theory is adiabatically connected to that
of the isotropic model. The low-energy spectrum
can be described by making use of the
approximate $\tilde{\mbox{G}}$-invariance.
In particular, the integrability of the resulting
model gives a description of the massive
quasiparticles which are organized in $\tilde{\mbox{G}}$-multiplets.

\item When the bare anisotropy is too large,
DSE cannot hold any longer, i.e. the energy
splitting of the $\tilde{\mbox{G}}$-multiplets due to
anisotropy becomes of the order of the gap
scale $m$:   the low-energy physics cannot
any longer be described in terms of a
single energy scale $m$, and the
situation is in general complex.

\item In between the symmetry enlarged
pockets lie self-dual manifolds, which are
the loci where anisotropy is maximal. A
simplification occurs, with in general a
decoupling of degrees of freedom, accompanied
with criticality. The quantum phase
transition that results -- which in the simplest
cases describes the transition
between the DSE phases -- is captured by
the minimal model with Hamiltonian (\ref{minmodel}).
Figure \ref{sketchcrossover} presents a qualitative picture
of the low-energy spectrum of this minimal model, along a path joining
two distinct symmetry enlarged phases. For clarity, one chooses a path
in parameter space along which the mass spectrum scales in a convenient
way: apart from possible phase transition points, the typical
mass scale is held constant.

\end{itemize}

%%%%%%%%%%%%%%%%%%%%%%%%%%%%%%%%%%%%%%%%%%%%%

\section{Fermionic models}
\label{secferm}

In this section, we come back to the
fermionic models (\ref{hamiltonian}) defined in
terms of $N$ chiral Dirac fermions
$\Psi_{a\L(\R)}$. We have shown, in the previous
section, that, irrespective
of the bare symmetry group H of the model,
there exists a finite number of possible
dualities that can be classified. To each of
them there corresponds a symmetry enlarged
phase, which we now characterize. There will result
an exhaustive list of DSE phases possibly
supported by models (\ref{hamiltonian}). For each
of those phases, the duality symmetries together with
the integrability yield precious informations on the low-energy physics.
As explained above, the maximal symmetry group supported
by those models is G = SU($N$) or G = SO$(2N)$, depending on the filling.
Here we analyze the general G = SU($N$) case which is relevant
to weakly-interacting fermions away from half-filling.
The special $N=4$ case, i.e. two-leg electronic ladders,
will be described in detail in section V.
The general half-filled case with G = SO$(2N)$ is much more
complicate to analyze and will be investigated elsewhere.

\subsection{Different classes away from half-filling}
\label{secincom}

For incommensurate filling,
there is no umklapp
process coupling spin and charge degrees
of freedom: $\Cal H = \Cal H_c + \Cal H_s$.
The charge fluctuations are decoupled and
described by a massless bosonic field $\Phi_c$
and its dual $\Theta_c$ field, with
Hamiltonian density:
\be
\Cal H_c = \frac{v_c}{2}\left( \frac{1}{K_c}(\p_x\Phi_c)^2
+  K_c(\p_x\Theta_c)^2\right),
\label{luttinger}
\ee
which belongs to the Luttinger liquid
universality class \cite{haldane,bookboso,giamarchi}. In
Eq. (\ref{luttinger}), $v_c$ and $K_c$ are the
charge velocity and Luttinger parameter,
respectively, that depend on the microscopic
details of the Hamiltonian and should be
regarded as phenomenological parameters.
For a commensurate filling, like one-electron per site,
additional umklapp processes may appear
depending only on the charge degrees of freedom
and the low-energy charge Hamiltonian (\ref{luttinger})
becomes a sine-Gordon model.
A charge gap might open for a sufficiently
strong value of the interaction which results in a Mott transition.
The non-trivial physics of model (\ref{hamiltonian}) corresponds to
the remaining ``spin'' degrees of freedom which are governed
by model (\ref{hamspinsector}) with
G = SU($N$).

According to the analysis of section \ref{sectdual},
we know that in the spin sector, there can only
be a finite number of symmetry enlarged phases,
that are classified by the SU($N$) $\Z_2$
gradings listed in Table \ref{tabZ2}. Due to DSE,
each of these phases has a finite extension
in the space of bare couplings, and its low-energy
sector is described by the representative
Hamiltonian (\ref{hamOmega}), which is labelled
by an involutive automorphism $\Omega$ of SU($N$).
This results in four  classes of
symmetry enlarged phases $\M_\Omega$, corresponding to
the trivial automorphism $\Omega=\idy$, and to
the three automorphisms belonging to the
classes A\I, A\II, A\III of Table I.
Recall that those are only the \emph{possible} dualities:
for them to be realized
in a given model defined by the symmetry breaking pattern
$\mbox{SU}(N)\downto \mbox{H}$, $H$ must contain in its center
an involutive element that gives rise to the duality.
As we will see, physical properties show
a strong distinction between symmetry
enlarged phases $\M_\Omega$ according to
whether they correspond to inner or
outer automorphisms $\Omega$ (here A\I and A\II correspond
to outer automorphisms, and A\III is inner).
While for inner dualities, the order is of the
density-wave type, phases associated to
outer dualities display off-diagonal order,
that comes along with superfluidity
in SU($N$) spin space. In addition, we will see that outer
dualities \emph{conjugate}, i.e. maps fermions
onto holes.
A representative for these three
classes for the SU($N$) group, in a suitable  basis for
the fermions $\Psi_{a\L}$ ($\Psi_{a\R}$
being invariant in the definition
of the dualities (\ref{twistdual})), is given by:

$\bullet$ \underline{A\I class} :
$\Psi_{a\L}\to \Omega(\Psi_{a\L})=\Psi_{a\L}^\dagger$.
The invariant subspace $\frg_\parallel$ is
$\fr{so}(N)$, with a basis defined by the subset of SU($N$)
generators $\Psi^\dagger_{a \L} T^i_{ab} \Psi_{b \L}^\low$,
$T^i$ being real and antisymmetric. Obviously, this
automorphism conjugates representations,
i.e. it maps the representation $\lambda$ onto
its charge conjugated partner $\lambda^*$.
The lattice symmetry
(the element of $\Cal C\big(\mbox{H}\big)_{\ind{inv}}$) giving
rise to $\Omega$ is charge conjugation:
$c_{j,a}^\low\to c_{j,a}^\dagger$.

$\bullet$ \underline{A\II class} :
$\Psi_{a \L}\to \Omega(\Psi_{a \L})=J_{ab}\Psi_{b\L}^\dagger$.
This duality exists only for even $N$ and the
matrix $J$ ($J$ being the Sp($N$) metric)
is given by $-i\sigma^2\times\idy_{N/2}$.
The invariant subspace $\frg_\parallel$ is $\fr{sp}(N)$,
with a basis defined by the subset of SU($N$)
generators $\Psi^\dagger_{a\L} T^i_{ab} \Psi_{b\L}^\low$
such that $JT^iJ=\;  ^t T^i$.
This automorphism also
conjugates representations.
The $N=2$ is very special since the duality corresponds to the identity
for the currents: Sp($2$) $\sim$ SU(2).
The lattice symmetry giving rise to $\Omega$ is an
``antisymmetric charge conjugation'':
$c_{j,a}^\low\to\sum_b J_{ab} c_{j,b}^\dagger$.
 
$\bullet$ \underline{A\III class} :
$\Psi_{a \L}\to\Omega(\Psi_{a \L})= \left(I_{p,q}\right)_{ab} \Psi_{b\L}$,
where $I_{p,q}$ ($0< p\leq q<N$) is the diagonal matrix
with $p$ entries 1 and $q$ entries -1. The
automorphism $\Omega$ belongs to Aut$_0$(G),
i.e. it is not an outer automorphism.
One has $\frg_\parallel=\fr{s}(\fr{u}(p)\oplus\fr{u}(q))$
as an invariant subspace.
The lattice symmetry that gives rise to this duality is:
$c_{j,a}^\low\to -c_{j,a}^\low$ for $a > p$. Dualities
in the A\III class are thus associated to a symmetry breaking
pattern $\mbox{G}\downto \mbox{H}$ that splits the $N$ fermionic
chains into two bunches of $p$ and $q=N-p$ chains.

%%%%%%%%%%%%%%%%%%%%%%%%%%%%%%%%%%%

\subsection{Physical properties of the fundamental phase $\M_\idy$}
\label{sectincommM0}

The physical properties of the symmetry enlarged
phases can be deduced from those of the
GN model (\ref{hamM0}), the representative
Hamiltonian corresponding to the phase $M_\idy$ labelled by
the trivial duality $\Omega=\idy$.

$\bullet$ \underline{Ground state and spectrum:}
Its exact spectrum is known from
integrability \cite{andrei79} and
consists in $N-1$ branches of quasiparticles
transforming in the Young tableau $\lambda_r$
with $r$ boxes and one column ($r=1...N-1$),
with masses $m_r=m\sin(\pi r/N)$. The quasiparticles
with label $r\neq 1,N-1$ can be seen as
boundstates of the ``fundamental'' quasiparticles
with $r=1$ and $r=N-1$. The states are labelled
by SU($N$) quantum numbers, i.e. by the eigenvalues
of the $N-1$ diagonal Cartan
generators $Q^\alpha_{\L}+Q^\alpha_{\R}$, $\alpha=1, \ldots,N-1$.

The order parameter of this spin-gapped phase
expresses simply in terms of the
SU($N)_1$ WZNW matrix field $\gwz$ which is defined by
\cite{kz}:
\be
\gwz_{ab}=e^{-i\sqrt{\frac{4\pi}{N}}\;  \Phi_c}\;
\Psi^\dagger_{a\L\vphantom{b}}\Psi^{\vphantom{\dagger}}_{b\R},
\label{defgsuN}
\ee
which is a pure spin field since
it commutes with all operators of the charge sector. This matrix field
identifies with the ``fundamental'' primary
operator, $\gwz=\Phi_{\lambda_{N-1},\lambda_1}$.
The order parameter reads then as follows:
\be
\vm{\Cal O_{\idy}}=\vm{\tr(\gwz)}\neq 0.
\label{OP0}
\ee
An order parameter on the lattice is readily
found by considering the $2k_\f$ oscillating part
of the total density operator
$n_i=\sum_{a=1}^N c^{\dagger}_{i,a}c^{\vphantom{\dagger}}_{i,a}$,
whose continuum limit is related to $\gwz$:
$n_{2k_\f}\sim e^{i\sqrt{\frac{4\pi}{N}}\Phi_c}\,\tr (\gwz)$.
This operator develops quasi-long-range order with a power-law
behavior for the equal-times two-point functions:
\be
\vm{n_{2k_\f}^\dagger(x)\,n_{2k_\f}(0)}\sim  x^{-\frac{2K_c}{N}}.
\label{CDW0}
\ee
If the charge sector develops a gap by some process
(for instance a pure charge umklapp term at
commensurate filling $1/N$ (one electron per site)
\cite{assaraf99}), the resulting phase is a Mott
insulator of the charge-density wave (CDW) type
with $\vm{n_{2k_\f} + \hc} \neq 0$.
This order may coexist with a $2k_\f$ bond-ordering
or spin-Peierls phase which is defined as:
${\cal O}_{\rm SP} = \sum_{j,a} e^{-2ik_{\f} ja_0^\low}\; c^{\dagger}_{j,a}
c^{\vphantom{\dagger}}_{j+1,a}  + \hc$
A lattice
order parameter involving only the ``spin''
degrees of freedom can also be found by considering
a generalized dimerization operator in SU($N$) space:
$\Cal O_{D}=\sum_{j,A}
 S^A_j S^A_{j+1}e^{-2ik_{\f} ja_0^\low}$,
where $ S^A_j=c^\dagger_{j,a}T^A_{ab}c^{\vphantom{\dagger}}_{j,b}$
is the SU($N$) spin lattice operator on
site $j$.
The latter operator has a continuum
limit involving different harmonics \cite{affleck,assaraf99}:
\be
 S^A_{j}\mathop{\sim}_{j=x/a_0}
\sum_{r=0}^{N-1} e^{2irk_\f x} \hat \alpha_r \Cal N^A_r(x),
\label{structfact}
\ee
where $\hat\alpha_r$ are vertex operators depending only on the charge boson --
if the charge sector is gapped, they can be replaced by non-universal amplitudes $\vm{\hat\alpha_r}$ 
-- and
$\Cal N^A_r=\tr (T^A_r \Phi_{\lambda_{N-r},\lambda_{r}})$
for $r\neq 0$, $T^A_r$ being the generators of
SU($N$) in the representation $\lambda_r$
($\mbox{dim}(\lambda_r)\times\mbox{dim}(\lambda_r)$ matrices).
The uniform spin density (i.e. $r=0$) is $\Cal N^A_0 = \J^A_{\L}+\J^A_{\R}$,
 with a universal amplitude $\hat\alpha_0 = 1/2\pi$.
Each of the harmonics $\Cal N^A_r$ transforms in the
adjoint representation of SU($N$) under the global
diagonal rotations.
When computed, the continuum
limit of $\Cal O_{D}$ shows that the SU($N$) dimerization operator
 develops a non-zero average value:
\be
\vm{\Cal O_{D}}\propto \vm{\tr (\gwz)  } \neq 0 .
\label{OPsp0}
\ee
We thus conclude, from the spin sector
point view,
that phase $\M_0$ corresponds to a
generalized dimerization phase, with an $N$-fold ground-state degeneracy for
a commensurate filling of one electron per site,
which is formed
from local SU($N$) spin-singlet condensation.

$\bullet$ \underline{Susceptibilities and rigidities:}
Due to the spin gap, the susceptibility at zero
temperature vanishes in all directions in
spin space. Indeed, coupling the model to some
field  $h^A\int dx\,(\J^A_{\L} + \J^A_{\R})$, one has
\be
\chi^{\vphantom{\dagger}}_{AB}\equiv\frac{\p^2 E[h]}{\p h^A\p h^B} =0,
\label{chi0}
\ee
where $E[h]$ is the energy of the system with the field $h$.
In contrast,
the quantity dual to susceptibility, namely the
rigidity that measures the response of the system
to twists in the boundary conditions
($c^{\vphantom{\dagger}}_{j+L/a_0,a} =
\Cal U_{ab}\;c_{j,b}^{\vphantom{\dagger}} $
with $\Cal U[\gamma]=e^{i\gamma^A T^A}$) for
a system on a ring of size $L$, is non-vanishing in all directions of spin space.
The rigidity is defined
as $\rho_{AB}^{\vphantom{\dagger}} \equiv
\frac{\p^2 E[\gamma]}{\p \gamma^A\p \gamma^B}$ and
one has $\rho_{AA}\neq 0$ for all $A=1,\ldots,N^2-1$.
This is easily seen by bosonizing model (\ref{hamM0}):
first one chooses a Cartan basis $(H^\alpha)$, $\alpha
=1, \ldots, N-1$
with the first element $H^1$ along the direction $A$,
and one introduces $N-1$ bosonic
fields $\varphi_{\alpha \L(\R)}$, $\alpha=1, \ldots, N-1$,
whose gradients represent the currents in the
Cartan directions: $\J^\alpha_{\L(\R)}
= \sqrt{4\pi}\p_x\varphi_{\alpha \L(\R)}$. The non-interacting
part of the Hamiltonian (\ref{hamM0}) is that of $N-1$
free bosons $\Cal H_{0s}= u\sum_\alpha
\left((\p_x\phi_\alpha)^2+ (\p_x\theta_\alpha)^2\right)/2$,
with $\phi_\alpha=\varphi_{\alpha\L}+\varphi_{\alpha\R}$
the total boson field and $\theta_\alpha=\varphi_{\alpha\L}
-\varphi_{\alpha\R}$ its dual field, while the interacting
part of model (\ref{hamM0}), $\Cal H_{\ind{int,s}}$, is built
from a sum of products
vertex operators $\Cal V_{\beta_\alpha,\bar\beta_\alpha}
= e^{i\beta_\alpha\varphi_{\alpha \L}
+ i\bar \beta_\alpha\varphi_{\alpha \R}}$. The
twist $\Cal U = e^{i\gamma H^1}$, along the
direction $A$ in the boundary conditions, results
in the following boundary conditions for the
bosonic fields: $\phi_\alpha(x+L)=\phi_\alpha(x)$
and $\theta_\alpha(x+L)=\theta_\alpha(x)
+(\gamma/\sqrt{\pi})\delta_{\alpha,1}$.
Now,
due to global (diagonal) SU($N$) invariance,
and in particular to the invariance under
the transformations generated by the
Cartan $H^\alpha$ that are represented by
a shift on the bosonic fields,
$\varphi_{\alpha\L(\R)}\to\varphi_{\alpha\L(\R)}\pm C_{\alpha}$,
$\Cal H_{\ind{int,s}}$ can involve only vertex
operators with $\beta_\alpha=\bar\beta_\alpha$,
i.e. it does not depend on the dual
field $\theta_\alpha$. It results that the twist
in the boundary condition can be absorbed
via the following canonical transformation
affecting the boson 1 only:
$\tilde\theta_1(x)=\theta_1(x)-\frac{\gamma}{L\sqrt{\pi}}\,x$,
$\tilde \phi_1(x)=\phi_1(x)$, yielding
\be
\rho_{AA}^{\vphantom{\dagger}}=\frac{1}{u\pi}\neq 0,
\label{rho0}
\ee
which signals the emergence of a finite rigidity.

$\bullet$ \underline{Spin excitations:}
Another quantity of interest characterizing the phase
is the dynamical spin structure factor $
S(q,\omega) = \int dxdt\,e^{-iqx-i\omega t}
\vm{S^A_{j+x/a_0}(t) S^A_{j}(0)}$.
Using the continuum description of the spin operators
(\ref{structfact}),
the Lehmann representation of
the zero-temperature spin
structure factor
involves the form factors
\be
\Cal F_r(\xi) = \vm{0|\Cal N^A_r |\xi},
\label{formfactor}
\ee
where $\ket{\xi}$ is an eigenstate of the
GN model (\ref{hamM0}). At low frequency, the behavior
of (\ref{formfactor}) is fixed by
one-particle states $\ket{\xi_{r'}}$ transforming in
the representation $\lambda_{r'}$. Since
the SU($N$) singlet representation does not
appear in the tensor product of the adjoint
by $\lambda_{r'}$, we deduce a strong
selection rule: $\Cal F_r(\xi_{r'}) = 0$ $\forall r,r'$.
The consequence is that the spin structure factor
displays no sharp peak structure at the quasiparticle
poles $q\approx 2rk_\f$ and $\omega\approx m_r$.
This situation is sometimes referred to
as \emph{incoherence} of the spin excitations.

%%%%%%%%%%%%%%%%%%%%%%%%%%%%%%%%%%

\subsection{Effect of dualities}

Acting with a duality $\Omega$ on the
GN model (\ref{hamM0}) yields the generalized
GN model (\ref{hamOmega}), which, in spite of
its simple connection to the original GN model,
displays qualitatively different physical
features.
We now deduce the non-trivial consequences of
the duality for the SU($N$) group on the low-energy physics
of the symmetry enlarged phases.

$\bullet$ \underline{Ground state:} For inner
dualities, using Eq. (\ref{primaryInn}),
we observe that the WZNW field $\gwz=\Phi_{\lambda_{N-1},\lambda_1}$
is mapped onto itself $\gwz\xrightarrow{\;\Omega\;}
U_{\lambda_{N-1}} \gwz$. As a result, the order parameter of this phase
is of CDW type, with a texture depending
on the precise form of the duality symmetry encoded in
the change of basis matrix
$U_{\lambda_{N-1}}$:
\be
\vm{\Cal O_\Omega} = \vm{\tr(U_{\lambda_{N-1}} \gwz)}\neq 0.
\ee
In a metallic phase, the $2k_\f$ component of the
corresponding lattice operator $n^\Omega_j=
\sum_{ab}c^\dagger_{j,a}(U_{\lambda_{N-1}})_{ba}c^{\vphantom{\dagger}}_{j,b}$
develops quasi-long-range correlations:
\be
\vm{n^\Omega_{2k_\f}(x)\,n^{\Omega\dagger}_{2k_\f}(0)}\sim x^{-2K_c/N} .
\ee
If some umklapp process opens a gap in
the charge sector, the ground state displays a
CDW
ordering characterized by $\vm{n^\Omega_{2k_\f}}\neq 0$.
The lattice generalized SU($N$) dimerization operator is now $\Cal O_D^\Omega
=\sum_{j,A}\epsilon^A\Cal S^A_j\Cal S^A_{j+1}e^{-2ik_\f ja_0^\low}$ with
$\epsilon^A=\pm1$
according to whether the direction $A$ belongs
to $\frg_\parallel$ or not.
Using the result (\ref{OPsp0}), we now conclude that
$\Cal O_D^\Omega
\sim\tr(U_{\lambda_{N-1}} \gwz )$, i.e. it
develops a non-zero expectation value indicating
a bond ordering.

In contrast, for outer dualities, using Eq. (\ref{primaryOut}), we see that
the WZNW field
is now changed into its dual, $\gwz\xrightarrow{\;\Omega\;}
U_{\lambda_{1}} \tilde \gwz$ where:
\be
\tilde \gwz_{ab} = e^{i\sqrt{\frac{4\pi}{N}} \;
\Theta_c}\;\Psi^\low_{a\L\vphantom{b}}\Psi^\low_{b\R}.
\label{defgtildesuN}
\ee
As a result, the order parameter of this phase
becomes:
\be
\vm{\Cal O_\Omega} = \vm{\tr(U_{\lambda_{1}} \tilde \gwz)}\neq 0.
\label{orderdual}
\ee
This translates
into \emph{non-diagonal}
ordering for the lattice fermions and the order parameter
becomes of a pairing type.
The uniform component
of the pairing density $\rho^\Omega_j=
\sum_{ab}c^\dagger_{j,a}(U_{\lambda_1})_{ba}c^\dagger_{j,b}$
develops critical correlations
\be
\vm{\rho^\Omega(x)\rho^\Omega(0)}\sim x^{-2/NK_c}.
\ee
In fact,
phases associated to outer dualities spontaneously
break a discrete symmetry of the original
lattice Hamiltonian (\ref{HamLatt}): the center $\Z_N$
of SU($N$), that consists of a discrete
gauge redefinition of the phase of the underlying
fermions $c_{j,a}\to e^{2i\pi/N}\,c_{j,a}$.
In the continuum limit, this transformation corresponds
to special phase factors on the Dirac fermions:
\be
\Z_N: \quad\Psi_{a \L(\R)}\to e^{2i\pi/N} \Psi_{a \L(\R)}.
\label{defZN}
\ee
For $N >2$,  $\rho^\Omega_j$ and $\Cal O_\Omega$
are clearly not invariant
under this discrete symmetry. The spontaneous breaking
of the center group symmetry
is thus responsible for the formation of the
spin gap of this phase.
Note that it is likewise sensible to interpret the formation of a gap
in phases associated to inner dualities as the spontaneous breaking
of a dual group $\tilde\Z_N$:
\be
\Psi_{a \L(\R)}\to e^{\pm 2i\pi/N} \Psi_{a \L(\R)},
\label{defZNtilde}
\ee
and this $\tilde \Z_N$ symmetry has no local representation on the original
lattice fermions.

$\bullet$ \underline{Spectrum:} The energy spectrum
of model (\ref{hamOmega}) is still
that of the GN model (\ref{hamM0}), but now the charges
labeling states, associated with the global symmetry
$\mbox{G}^{I\!R}=\big(\tilde{\mbox{G}}_{\L}
\times\mbox{G}_{\R}\big)_{\ind{diag}}$, are
the $N-1$ Cartan generators $H^\alpha_\Omega=
Q^\alpha_{\R}+\tilde Q^\alpha_{\L}$, $\alpha=1, \ldots, N-1$,
with $\tilde Q^\alpha_{\L}=\epsilon^\alpha Q^\alpha_{\L}$,
$\epsilon^\alpha =\pm1$
according to whether the direction $\alpha$ belongs
to $\frg_\parallel$ or not. It is possible to show that
for inner dualities, one can always find a
Cartan basis untouched by $\Omega$, i.e.
with $\epsilon^\alpha=1$ $\forall \alpha$, and
we will work in such a basis. It results that
states can be labelled by Noether
charges $J^\alpha_0$, see Eq. (\ref{noether}). For
outer dualities though, it is not the case
(for otherwise any representation $\lambda$
would be mapped to itself). Instead, there is at
least one direction $\alpha$ in Cartan space
such that $H^\alpha_\Omega = J^\alpha_1$, i.e.,
the Noether current. We conclude that in the diagram (\ref{schemaresto}),
there is some sort of transmutation from Noether
charges to Noether currents when going from UV to IR.

$\bullet$ \underline{Susceptibilities and rigidities:}
Following again the same line of reasoning
that lead to the results (\ref{chi0},\ref{rho0}), one sees
that dualities, in spite of the presence of the
 mass gap in every sector of the spectrum, results
in the existence of directions in $\fr{su}(N)$
with \emph{non-vanishing} susceptibilities
$\chi^{\vphantom{\dagger}}_{AA}$, and vanishing
rigidity $\rho^{\vphantom{\dagger}}_{AA}$, namely
those ``transmuted'' directions for which
$\epsilon^A = -1$. This is obviously consistent with the fact 
the associated quantum number along the direction $A$ is
a current and not a charge. 
If the duality is outer, \emph{any} Cartan basis (that is, any choice of
the quantization axis for the quantum numbers labeling the states) will be touched
by the duality, i.e. it will contain at least one direction
with $\epsilon^A=-1$, and vanishing rigidity in that direction.
It turns out that this is connected to an 
observable physical effect, that allows to distinguish between inner
and outer dual phases: as shown in Ref.\onlinecite{lecheminant05},
 the breaking of the $\Z_N$ symmetry (\ref{defZN}) in outer phases 
 leads to a "confinement" of the current, elementary gapless excitations
 \footnote{This effect appears at incommensurate filling, 
 i.e. with no gap in the charge sector.}
 carrying currents quantized in units of $Ne$. 
 For a finite system put on a rotating ring,
 this results in a 
 modified periodicity
 of the groundstate energy
 as a function of tangential velocity \cite{lecheminant05,seidel}.

$\bullet$ \underline{Spin excitations:}
For inner dualities, the conclusions drawn
in the case of the trivial duality $\Omega=\idy$ are
not affected: spin excitations are still incoherent. However,
for outer dualities, the spin density $\Cal N^A_r$
at wave vector $q=2rk_\f$ in Eq. (\ref{structfact}) is
mapped onto $\Omega(\Cal N^A_r)=\tr(T^A_r \,U_{\lambda_r}
\Phi_{\lambda_{r},\lambda_{r}})$, and owing to
the tensor product $\lambda_r\otimes\lambda_r=\lambda_{2r}\oplus
 \hdots$ (indices are understood modulo $N$), the form
factor of Eq. (\ref{formfactor}) $\Cal F_{r}(\xi_{N-2r})$ is
non-vanishing. Hence, the dynamical spin structure factor displays
a sharp peak at $q=2rk_\f$ and $\omega=m_r$: spin
excitations in a charge-gapped  system are now \emph{coherent}
\footnote{This conclusion only holds in a charge gapped phase. If charge excitations are gapless,
then the fluctuations in the charge sector (because of the presence of the operators $\hat\alpha_r$ in the spin operator, see Eq.(\ref{structfact})) will wash out
the coherence in the spin-spin correlation function \cite{pellobianco,esslerkonik}.}. 
Outer dualities thus have a
simple spectroscopic signature.

\begin{table}[h]
\begin{center}
\begin{tabular}{|c|c|c|c|}
\hline
Duality $\Omega$ \& phase &
order parameter
& inn/out
& \begin{tabular}{c}lattice symmetry  \\ giving rise to $\Omega$\end{tabular}
\\
\hline
\begin{tabular}{c}
   Trivial
   \\
   SU(N) CDW-SP
\end{tabular}
&
$\begin{array}{c}
c^\dagger_{j,a}c_{j,a}^\low \;e^{-2ik_\f ja_0}\\
c^\dagger_{j,a}c_{j+1,a}^\low \;e^{-2ik_\f ja_0}
\end{array}
$
&
inner
&
identity
\\
\hline
\begin{tabular}{c}
   A\I
   \\
   Symmetric pairing
\end{tabular}
&
$c^\low_{j,a} c^\low_{j+1,a} $
&
outer
&
   $c_{j,a}^\low\to c_{j,a}^\dagger$
\\
\hline
\begin{tabular}{c}
   A\II
   \\
   Antisymmetric pairing
\end{tabular}
&
$c^\low_{j,a} J_{ab} c^\low_{j,b}$
&
outer
&
\begin{tabular}{c}
   $c_{j,a}^\low\to J_{ab}c_{j,b}^\dagger$
   \\
   (\scriptsize $J=-i\sigma_2\otimes \idy_{N/2}$)
\end{tabular}
\\
\hline
\begin{tabular}{c}
   A\III
   \\
   ``Split chains''
   \drawbunchesmall
\end{tabular}
&
$
\begin{array}{c}
\big(\frac{1}{p}\sum_{a\leq p} - \frac{1}{q}\sum_{a>p} \big)
c^\dagger_{j,a} c^\low_{j,a} e^{-2ik_\f ja_0}\\
\big(\frac{1}{p}\sum_{a\leq p} - \frac{1}{q}\sum_{a>p} \big)
c^\dagger_{j,a} c^\low_{j+1,a} e^{-2ik_\f ja_0}
\end{array}
$
&
inner
&
\begin{tabular}{c}
   $c_{j,a}^\low\to -c_{j,a}^\low$
   \\
   (\scriptsize $a > p$)
\end{tabular}
\\
\hline
\end{tabular}
\caption{
\label{recapOmega}
The different classes of possible symmetry enlarged phases
supported by the $N$-component degenerate fermions
 in the absence of spin-charge coupling
(incommensurate case). For each duality $\Omega$, we give the corresponding
lattice order parameter, the nature (inner or outer) of the duality, and the
lattice symmetry giving rise to $\Omega$.}
\end{center}
\end{table}

In Table \ref{recapOmega}, we present a summary
of the different possible symmetry enlarged
phases for the
$N$-component degenerate fermions close to the SU$(N)_1$ critical point,
considered away
from half-filling.

\subsection{Cross-overs and quantum phase transitions}

In this section we investigate self-dual points and their vicinity.
We have seen in section \ref{secQPTgen} that those are points
where the competition between two different symmetry enlarged
phases is maximal, and can possibly lead to a quantum phase transition.
For each pair of dualities $\Omega_1$ and $\Omega_2$,
the region lying between phases $\M_{\Omega_1}$ and $\M_{\Omega_2}$
associated to them  is best described in terms of the simplest model
(\ref{minmodel}). By acting with the duality $\Omega_1$,
one can reduce the family of those interpolating models that need
to be investigated: it is enough to consider the models interpolating
between $\M_\idy$ and $\M_\Omega$, where $\Omega\neq\idy$ is any
non-trivial duality:
\be
{\Cal H}_{\idy-\Omega}
= \Cal H_0 + g_\parallel\,\J_{\R\parallel}^a\J_{\L\parallel}^a
 + g_\perp\,\J_{\R\perp}^b\J_{\L\perp}^b.
\label{minOmega}
\ee
At the self-dual point ($g_{\perp} = 0$),
the interaction in the parallel sector gives rise
to a gap for $g_\parallel > 0$ (this is the case 
we will consider in the following
\footnote{The case $g_\parallel < 0$, for which the interaction
at the self-dual point is marginally irrelevant, will
not be analyzed here. It corresponds to situations
where additional criticality is present at the self-dual point: criticality
of the multicritical point extends in parameter space. We will present in detail an example
of this in section \ref{secexample}.
}):
the spectrum in this sector is this of the
$G_{\parallel \iota}$ GN model, where $\iota$
is the embedding index of the embedding
$G_\parallel\subset G$ associated to $\Omega$.
There are, however, other degrees of freedom in general,
that are not affected by the interaction: this decoupled sector
is described by the coset model
$\Cal A = \frac{\mbox{\scriptsize SU}(N)_1}
{\mbox{\scriptsize G}_{\parallel \iota}}$ (see Eq. (\ref{defcosetA})).
If model $\Cal A$ is non-trivial, i.e. with central
charge $c_{\Cal A}^\low \neq 0$, this is an indication that
there is room for a continuous phase transition, in the universality class
of the CFT $\Cal A$. If it is trivial,
a phase transition might occur as we will see.
The inner or outer nature of the duality will
have different striking effects.

$\bullet$ \underline{A\I transition:}
The Dynkin index for the embedding SO$(N)\subset\mbox{SU}(N)$ is $\iota=2$,
so that model $\Cal A$ is trivial: $c_{\Cal A}^\low = 0$.
The only degrees of freedom at the self-dual point are thus
those of the parallel sector, which are governed by the following Hamiltonian:
\be
{\Cal H}^{*}_{\Omega} =
\Cal H_0 + \lambda \J_{\parallel\L}^a\J_{\parallel\R}^a ,
\ee
where $\J_{\parallel}^a$ are SO$(N)_2$ currents.
\footnote{We consider
here the generic case $N > 2$ since
the $N=2$ case is very special. As discussed in Ref. \onlinecite{phletotsuka},
the resulting transition for $N=2$ is described
by a SU(2) self-dual sine-Gordon model which displays an U(1)
quantum criticality \cite{phleSDSG}.}
This is the $k=2$
SO($N$) GN model which is fully gapped for
$\lambda > 0$.%
\footnote{A subtlety arises here from the fact that the conformal embedding
 $\mbox{SO}(N)_2\subset \mbox{SU}(N)_1$, yields a non-diagonal invariant for
the partition function in the $\mbox{SO}(N)$ variables. Namely, some
$\mbox{SO}(N)_2$ states have to be projected out. Specifically, when $N=2n$ is
even, the $\mbox{SO}(2n)_2$ primary operators constituting
the non-interacting
spectrum (that are necessary to reconstruct the $\mbox{SU}(2n)_1$ Hilbert
space), together with their degeneracy, are given by:
$1\times\idy,
\big\{2\times\Phi_{\lambda_j,\lambda_j}\big\},
1\times \Phi_{\lambda_s^++\lambda_s^-,\lambda_s^++\lambda_s^-},
1\times\Phi_{2\lambda_s^\pm,2\lambda_s^\pm},
1\times\Phi_{2\lambda_s^\pm,2\lambda_s^\mp},
1\times\Phi_{2\lambda_1,2\lambda_1}
$.
Here the integer $j$ runs from 1 to $n-2$, and $\lambda_j$ is the highest
weight of the representations of $\mbox{SO}(N)$, while the spinorial
representations (which do not appear in the
spectrum) have highest weight $\lambda_s^\pm$. When $N=2n+1$ is odd, the
spectrum is given by:
$
1\times\idy,
\big\{2\times\Phi_{\lambda_j,\lambda_j}\big\},
1\times\Phi_{2\lambda_s,2\lambda_s},
1\times\Phi_{2\lambda_1,2\lambda_1}
$, the integer $j$ running from 1 to $n-1$.
We make the reasonable hypothesis that this modified $\mbox{SO}(N)$
GN
model at level 2 is still fully gapped. The study of its spectrum goes beyond
the scope of this work.}
In particular, it has been shown in Ref.  \onlinecite{phletotsuka}
that this model is related to the Andrei-Destri \cite{andrei} integrable model
which is exactly solvable by means of the Bethe ansatz
approach \cite{andrei,reshtekin}.
Therefore,  no criticality
emerges at the self-dual point. However, recalling that $\Omega$
is outer and conjugates representations, we can show that the order
parameters on both sides
cannot coexist: while $\tr (\gwz)$ is invariant under the center of SU(N),
that is under the $\Z_N$ transformations (\ref{defZN}), it is not the case
for the off-diagonal order parameter $\tr(U_{\lambda_1}\tilde\gwz)$
of Eq. (\ref{orderdual}) for $N > 2$.
The $\Z_N$ symmetry is thus spontaneously broken in phase $\M_\Omega$.
The opposite is true for the dual symmetry $\tilde \Z_N$ (\ref{defZNtilde}),
which
leaves invariant $\tilde \gwz$ but not $\gwz$, so that it is
spontaneously broken in phase $\M_{\idy}$.
We conclude to the occurrence of a first-order transition.

$\bullet$ \underline{A\II transition:}
The embedding index is one in this case and
one has the conformal
embedding \cite{altschuler89}
$\mbox{SU}(2n)_1\sim \mbox{Sp}(2n)_1 \times \Z_n$,
(recall that $N=2n$ is even in this case)
where the last piece, corresponding
to $\Cal A$, stands for the parafermionic minimal model with
$\Z_n$  symmetry ($n \ge 2$)
and central charge $c_{\Cal A}^\low = 2(n-1)/(n+2)$ which is a
generalization of the
Ising $\Z_2$  critical theory \cite{zamolofateev}.
This transition was studied at
length in Ref. \onlinecite{lecheminant05,lecheminant05b}. In particular,
it was shown that for $N<4$,
this criticality is stable, i.e. no relevant operator in
the $\Z_n$  sector is allowed
by symmetries that would spoil criticality. Once again, the
quantum phase transition is associated
to the spontaneous breaking of the $\Z_n$ symmetry --
with respect to the A\I case,
a $\Z_2$  subgroup of the center of SU$(2n)$ survives
that corresponds to a redefinition
of the (lattice) fermion sign ($c_{i,a} \rightarrow - c_{i,a}$).

$\bullet$ \underline{A\III transition:}
The embedding index is one, and the coset model $\Cal A$ is trivial,
with $c_{\cal A}^\low=0$. Recall that the automorphism $\Omega$ is defined
by splitting the set of $N$ fermionic chains into two clusters of size $p$ and
$q$ with $p+q=N$.
The affine
symmetry $\fr{g}_\parallel^\low$ decomposes in three simple factors:
 $\fr{su}(p)_1\oplus\fr{su}(q)_1\oplus\fr{u}(1)$ (see
Table I). The first two factors
are generated by the affine currents $\J_{\L(\R)}^{(p)}$ and
$\J_{\L(\R)}^{(q)}$, that
are respectively the uniform components of the lattice spin operators
$\sum_{a,b\leq p}c^\dagger_{j,a} T_{ab}^{(p)}c^\low_{j,b}$ and
$\sum_{p<a,b}c^\dagger_{j,a} T_{ab}^{(q)}c^\low_{j,b}$,
where $T^{(p)}$ (respectively $T^{(q)}$) are traceless Hermitian
matrices with non-zero entries only in the upper-left $p\times p$ block
(lower-right $q\times q$ block respectively).
The affine current $\J^0$, generating the U(1) piece, is the uniform component
of the relative charge of the two chain clusters
$Q_f = \frac{1}{\sqrt{Npq}}
\left(q\sum_{j,a\leq p}c^\dagger_{j,a}c^\low_{j,a}
-p \sum_{j,a>p}c^\dagger_{j,a}c^\low_{j,a}\right)$.
This current can be represented by chiral bosonic fields
$\phi_{f\L(\R)}$ according to
$\J^0_{\L(\R)} = \sqrt{4\pi}\,\p_x\phi_{f\L(\R)}$.
The order parameter in phase $\M_{\idy}$ can then be expressed in terms
of the WZNW matrix fields $\gwz^{(p)}$, $\gwz^{(q)}$
of $\mbox{SU}(p)_1$ and  $\mbox{SU}(q)_1$:
\be
\tr(\gwz) = e^{i\sqrt{\frac{4\pi q}{Np}}\phi_f}\,\tr(\gwz^{(p)} )
+ e^{-i\sqrt{\frac{4\pi p}{Nq}}\phi_f}\,\tr(\gwz^{(q)}),
\label{wzsupsuq}
\ee
with $\phi_f = \phi_{f \L} + \phi_{f \R}$.
A very similar expression holds for the order parameter in phase $\M_\Omega$:
$\tr(\Omega(\gwz)) = e^{i\sqrt{\frac{4\pi q}{Np}}\phi_f}\,\tr(\gwz^{(p)} )
- e^{-i\sqrt{\frac{4\pi p}{Nq}}\phi_f}\,\tr(\gwz^{(q)})$.
Generically, the two order parameters have the same symmetry
and one cannot exclude coexistence.
The self-dual model is described by the following
 Hamiltonian density:
\be
\Cal H^{*}_{\Omega}
 = \Cal H_f + \Cal H_{(p)}+ \Cal H_{(q)} ,
\label{selfdualA3}
\ee
where $\Cal H_{(p)}$ is the
SU$(p)$ GN model at level one (similarly for $\Cal H_{(q)}$),
so that the corresponding parallel degrees of freedom are fully gapped.
The boson $\phi_f$ decouples and is governed by:
\be
 \Cal H_f = \frac{v_\f}{2}\left((\p_x\phi_f)^2
+ (\p_x\theta_f)^2\right) + g_f^\low\,\J^0_\L\J^0_\R
= \frac{u_f}{2}\left(\frac{1}{K_f}(\p_x\phi_f)^2
+ K_f(\p_x\theta_f)^2\right),
\label{hamphif}
\ee
which is nothing but the Luttinger Hamiltonian, with
velocity $u_f=\sqrt{v_\f^2-4\pi^2 g_f^2}$ and Luttinger
parameter $K_f=\sqrt{v_\f-2\pi g_f}/\sqrt{v_\f+2\pi g_f}$.
This approach predicts a central charge
$c=1$ on the self-dual manifold.
Of course, naively irrelevant operators (i.e. operators that are
irrelevant with respect to the fully unperturbed SU($N)_1$ WZNW model) could
be present that would spoil this criticality in the $\phi_f$ sector.
The situation is complex in
general, and we will only investigate here the incommensurate case,
thus ignoring umklapp oscillating terms.
Of course, since we start with a Hamiltonian local in terms
of the lattice fermions, the perturbing  operators can be built
out of the fermions $\Psi_{\L(\R)}$. The only
$\mbox{SU}(p)\times\mbox{SU}(q)\times\mbox{U}(1)$
candidates are of the form
$\Cal O_k =
\big(\prod_{j=1}^k\Psi^\dagger_{a_j^\low \L}\Psi^\low_{a_j^\low \R}\big) \;
\big(\prod_{j=1}^k\Psi^\dagger_{b_j^\low \R}\Psi^\low_{b_j^\low \L}\big)
+ \hc$,
where antisymmetrization is implied in the indices
$a_j^\low\leq p$ and $b_j^\low > p$, and thus necessarily $k\leq p,q$.
Moreover, the self-dual symmetry forces $k$ to be an even integer.
These operators can be expressed in terms of the SU$(p)_1$
and SU$(q)_1$ primary operators:
$\Cal O_k \sim \tr(\Phi^{(p)}_{\lambda_k^*,\lambda_k})\,
\tr(\Phi^{(q)}_{\lambda_k,\lambda_k^*}) \;
e^{ik\sqrt{\frac{4\pi N}{pq}}\phi_f} + \hc$,
where in obvious notations $\Phi_{\lambda_k^*,\lambda_k}^{(p)}$
denotes the SU$(p)_1$ primary operator in the ``$p$'' sector.
On the self-dual manifold, it is legitimate to average out the gapped
degrees of freedom, thus yielding the following effective
perturbing operator in the $\phi_f$ sector:
\be
\bar {\Cal O}_k \sim g\;
\cos\bigg(\sqrt{\frac{4\pi Nk^2}{pq}}\phi_f\bigg),
\label{pertA3}
\ee
where $g\propto\vm{\tr(\Phi^{(p)}_{\lambda_k^*,\lambda_k})}_{GN}
\vm{\tr(\Phi^{(q)}_{\lambda_k,\lambda_k^*})}_{GN}
$ (the average is taken in the SU$(p)$ and SU$(q)$ GN model) is a
non-universal prefactor.
The dominant perturbing operator is thus $\bar {\Cal O}_2$
with a scaling dimension (with respect to the Luttinger liquid fixed point)
$\Delta_2 = 4NK_f/pq$. We thus see that depending on the value of $K_f$ the
model exhibits different behaviors on the self-dual manifold: when $K_f<pq/2N$,
the operator $\bar{\cal O}_2$ is relevant, and spoils criticality:
in this case, model (\ref{selfdualA3}) describes a smooth cross-over
between phases $\M_{\idy}$ and $\M_\Omega$. On the other hand,
if $K_f>pq/2N$, there is a true continuous phase transition,
with a $c=1$ criticality emerging for the degrees of freedom describing
the relative charge on the two clusters of fermionic chains.
In the latter case,
 it is instructive to consider the Hamiltonian describing the vicinity
of the self-dual manifold. Writing the currents $\J_{\perp}$ in terms of the
$\mbox{SU}(p)_1\times\mbox{SU}(q)_1\times\mbox{U}(1)$ variables yields:
\be
\Cal H_\Omega = \Cal H_\Omega^*  + g_\perp\,\tr\left(\gwz^{(p)}\right)
\tr\left(\gwz^{(q) \dagger}\right) \, e^{ i\sqrt{\frac{4\pi N}{pq}}\,\phi_f}
+ \hc
\ee
The perturbing operator is nothing but $\Cal O_1$.
Averaging out the gapped degrees of freedom (this is
legitimate in the limit of
small $g_\perp$),
one concludes that the low-energy degrees of freedom close to the
self-dual manifold, described by the bosonic
field $\phi_f$, are again governed by a sine-Gordon model.
The scaling dimension of its
cosine term $\Delta_1 = NK_f/pq$ so that a gap
opens if $K_f<2pq/N$ as soon as $g_\perp\neq 0$.
If $K_f>2pq/N$, the perturbation is irrelevant and
one can only conclude that the $c=1$
criticality extends on both sides of the self-dual manifold. Though this
criticality is doomed to disappear when one departs too far from the self-dual
manifold, our approach does not allow to put a bound to this criticality in
parameter space.

\begin{table}
\begin{center}
\begin{tabular}{|c|c|c|}
\hline
duality type & criticality on the self-dual
manifold & critical degrees of freedom \\
\hline
\begin{tabular}{c}
A\I \vspace*{-0.2cm}  \\ (symmetric pairing) \vspace*{0.2cm}
\end{tabular}
&
no criticality (first order transition)
& $-$ \\
\hline
\begin{tabular}{c}
A\II \vspace*{-0.2cm}  \\ (antisymmetric pairing) \vspace*{0.2cm}
\end{tabular}
&
\begin{tabular}{ll}
   $N=2n \leq 8$: & $\Z_{n}$ parafermions
   \\
   $N=2n > 8$: & first order transition
\end{tabular}
&
\begin{tabular}{c}
   associated to the symmetry $c_{i,a}\to e^{2i\pi/N} c_{i,a}$
   \\
   $-$
\end{tabular}
\\
\hline
\begin{tabular}{c}
A\III \vspace*{-0.2cm}  \\ ($p$ and $q$ bunches) \vspace*{0.2cm}
\end{tabular}
&
\begin{tabular}{ll}
   $K_f<\frac{pq}{2N}$: & smooth crossover
   \\
   $\frac{pq}{2N}<K_f<\frac{2pq}{N}$: \quad& $c=1$ transition
   \\
   $K_f>\frac{2pq}{N}$: & $c=1$ pocket
\end{tabular}
&
\begin{tabular}{c}
   $-$
   \\
   relative charge of the two bunches of chains
   \\
   relative charge of the two bunches of chains
\end{tabular}
 \\
\hline
\end{tabular}
\caption{
\label{QPTincomm}
Different classes of quantum phases transitions between the symmetry enlarged
phases, away from half-filling.}
\end{center}
\end{table}

To restate our conclusions for the A\III class, we find that the behavior close to
the self-dual manifold depends on the value of the Luttinger parameter $K_f$,
that can vary continuously. If $K_f<pq/2N$, one has a smooth crossover from
phase $\M_\idy$ to phase $\M_\Omega$. If $pq/2N<K_f<2pq/N$, there is a quantum
phase transition with $c=1$, the self-dual model being a Luttinger liquid for
the relative charge between the two clusters of chains. If $K_f>2pq/N$, this
criticality acquires a finite extension around the self-dual manifold.

Table \ref{QPTincomm} presents a summary of the quantum phase transitions
that occur between the different classes for general fermionic models
away from half-filling.
%%%%%%%%%%%%%%%%%%%%%%%%%%%%%%%%%%%%%%%%%%%%%
%%%%%%%%%%%%%%%%%%%%%%%%%%%%%%%%%%%%%%%%%%%%%

\section{Dualities at work: the example of the two-leg electronic ladder}
\label{secexample}

In this section we will apply the duality approach,
presented in the preceding sections, to the study
of a specific example.
The novelty of this approach stems from the fact
the different possible symmetry
enlarged phases can be exhausted by means of the algebraic properties
of the groups G and H, instead of solving numerically the one-loop RG flow
and scanning the different phases.
In addition, once a phase is identified and characterized,
there is a systematic way to obtain the physics
of the other duality-related phases.
Finally, the duality approach also offers the possibility to
study directly phase transitions between the
different symmetry enlarged phases,
by investigating the self-dual manifolds.
We now turn to a careful examination
of a particular example, namely a generalized two-leg electronic
ladder.

%%%%%%%%%%%%%%%%%%%%%%%%%%%%%%%%%%%%%%%%%%%%%

\subsection{Lattice model}

Here we apply the ideas exposed in the preceding sections to the model
of two-leg Hubbard ladder with
Hamiltonian:
\be
\Cal H = -\sum_{i}\sum_{\ell\ell'\sigma\sigma'} t_{\ell\sigma,\ell'\sigma'}\;
\left(c^\dagger_{i,\ell\sigma}c^{\vphantom{\dagger}}_{i+1,\ell'\sigma'}
+ \hc \right)
+\sum_{i}\sum_{\ell_j\sigma_j}U_{\{\ell_j\}\{\sigma_j\}}\;
c^\dagger_{i,\ell_1\sigma_1}c^\dagger_{i,\ell_2\sigma_2}
c^{\vphantom{\dagger}}_{i,\ell_3\sigma_3}
c^{\vphantom{\dagger}}_{i,\ell_4\sigma_4} ,
\label{Hlatt2leg}
\ee
where $c_{i,\ell\sigma}^\dagger$ creates an
electron with spin $\sigma=\uparrow,\down$ on chain $\ell=1,2$, at site $i$.
The explicit form
of the hopping parameters $t_{\ell\sigma,\ell'\sigma'}$ and
on-site couplings $U_{\{\ell_j\}\{\sigma_j\}}$
depends on the precise model under consideration, that
can be specified by fixing the bare
physical symmetry group H.

For incommensurate filling, there is a spin-charge
separation and, as already discussed in section II,
the relevant maximal symmetry group G supported
by the two-leg ladder (\ref{Hlatt2leg}) is
G=SU(4) that mixes the four
different local states $\big\{c^\dagger_{i,\ell\sigma}\ket{0}\big\}_{\ell
=1,2;\sigma=\uparrow,\down}$ (here $\ket{0}$ is the vacuum state).
We now need to specify the bare symmetry
group H. Here, we have chosen to present the
example of a model that retains three
basic symmetries: first, a natural SU(2) symmetry
acting in the spin sector; second, a $\Z_2$
symmetry that exchanges the two chains of
the ladder; and finally, a U(1) orbital symmetry
generated by $N_1-N_2$, where $N_\ell=
\sum_{i\sigma} c^\dagger_{i,\ell\sigma}
c^{\vphantom{\dagger}}_{i,\ell\sigma}$ is the
relative charge on the two chains. This leads to the
following symmetry breaking pattern ignoring the
U(1) charge symmetry:
\be
\mbox{G}\downto \mbox{H}=\mbox{SU(2)}_{\ind{spin}}
\times\mbox{U(1)}_{\ind{orb}}\times\Z_2 \times
\mbox{H}_{\ind{discrete}},
\label{symbreakpatt}
\ee
where $\mbox{H}_{\ind{discrete}}$ accounts for
the remaining discrete symmetries of model (\ref{Hlatt2leg}).
As discussed later, without affecting our conclusions
regarding the number and labeling of phases,
the constraint of the U(1) orbital symmetry can
also be relaxed: in this case, and in the
generic case (i.e. with a non-vanishing interchain
hopping that is permitted in this case), we
recover in the continuum limit an
\emph{effective} $\tilde{\mbox{U}}$(1) symmetry group,
so that the symmetry breaking pattern (\ref{symbreakpatt})
still holds.
As a consequence, the structure of the phase
diagram, that relies on the pattern (\ref{symbreakpatt}),
will be the same.

Given the symmetry H, we can
construct the Hubbard-like Hamiltonian
exhibiting this symmetry (for simplicity,
we consider only on-site couplings) \cite{arovas}.
The hopping amplitudes are constrained to
take the value $t_{\ell\ell'\sigma\sigma'}
=t\,\delta_{\ell\ell'}\delta_{\sigma\sigma'}$.
As for the terms quartic in fermions, they are
obtained by considering the general
two-fermion states at site $i$,
$c^\dagger_{i,\ell\sigma}c^\dagger_{i,\ell'\sigma'}\ket{0}$.
These states decompose under the group H
as $(1;0)\oplus (0;0)\oplus (0;\pm 1)$,
where $(S;L)$ denotes the multiplet of spin $S$
that carries orbital U(1) charge $L$. Denoting
the 6 local two-particle states
by $\ket{S,S^z;L}$, they read explicitly as follows:
\bea
\ket{1,1;0}&=&c^\dagger_{1\upa}c^\dagger_{2\upa}\ket{0}\nonumber\\
\ket{1,-1;0}&=&c^\dagger_{1\down}c^\dagger_{2\down}\ket{0}\nonumber\\
\ket{1,0;0}&=&
\frac{1}{\sqrt{2}}\, \left(c^\dagger_{1\upa}c^\dagger_{2\down}
+c^\dagger_{1\down}c^\dagger_{2\upa}\right)\ket{0}\nonumber\\
\ket{0;1}&=&c^\dagger_{1\upa}c^\dagger_{1\down}\ket{0}\nonumber\\
\ket{0;-1}&=&c^\dagger_{2\upa}c^\dagger_{2\down}\ket{0}\nonumber\\
\ket{0;0}&=&
\frac{1}{\sqrt{2}}\, \left(c^\dagger_{1\upa}c^\dagger_{2\down}
-c^\dagger_{1\down}c^\dagger_{2\upa}\right)\ket{0} ,
\eea
and they allow for three different
invariants under the group H,
namely $\sum_{S^z=0,\pm 1}\ket{1,S^z;0}\bra{1,S^z;0}$,
$\ket{0,0;1}\bra{0,0;1} + \ket{0,0;-1}\bra{0,0;-1}$,
and $\ket{0,0;0}\bra{0,0;0}$. In this respect, we choose to
parametrize the interacting-part of
the Hamiltonian (\ref{Hlatt2leg}) as:
\be
H_{\ind{int}} = \sum_ i \Big[ \;\frac{U}{2}\!\! \sum_{(\ell\sigma)\neq(\ell'\sigma')}
n_{i,\ell\sigma}n_{i,\ell'\sigma'} + J_H \;\vec{S}_{i,1}\cdot\vec{S}_{i,2}
+ J_t \;\left( T^z_i  \right)^2
\Big] ,
\label{hundmodel}
\ee
 where
$n_{i,\ell\sigma}=c^\dagger_{i,\ell\sigma}c^{\vphantom{\dagger}}_{i,\ell\sigma}$
is the local electronic density for
species $(\ell\sigma)$,
$\vec{S}_{i,\ell} =\frac{1}{2}
\sum_{\sigma\sigma'}
c^\dagger_{i,\ell\sigma}\vec{\sigma}_{\sigma\sigma'}
c^{\vphantom{\dagger}}_{i,\ell\sigma'} $ is the
spin operator on chain $\ell$,
and $T^z_i=\frac{1}{2} \sum_{\sigma}\left( n_{i,1\sigma}- n_{i,2\sigma}\right)$
is the difference of electronic densities
on the two chains, that generates the U(1) orbital symmetry.

The resulting Hamiltonian depends on three
microscopic couplings: an (overall)
Coulombic interaction $U$, a Hund coupling $J_H$,
and an ``orbital crystal field anisotropy'' $J_t$.
Playing with these three parameters allows to recover
several limiting cases that have already been studied:
when $J_H=J_t=0$, one recovers the SU(4) Hubbard model,
that has been extensively analyzed in
recent years \cite{assaraf99,assaraf04,solyom,ueda,capponi}.
The case where only Hund coupling and
Coulomb interaction are present ($J_t=0$) has been
studied in Refs. \onlinecite{lee} and \onlinecite{fabrizio}, where it has been
shown that the Hund coupling can stabilize
a $d$-wave superconducting instability.
When $J_t=3 J_H/4$, the symmetry is
promoted to SU(2)$_{\ind{spin}}\times$SU(2)$_{\ind{orb}}$,
and one recovers the so-called spin-orbital
model \cite{yamashita,khomskii,mila,aza4spin,itoi00}. When $J_t=J_H/4$,
the symmetry is enlarged to SO(5)$\times\Z_2$, and
this model has been studied in the
context of cold-atoms
physics \cite{wu,lecheminant05,lecheminant05b,capponi,capponi2007,tsvelik06}.
Finally, as it will be discussed later,
we note that two-leg electronic ladders with
a $t_{\perp}$ hopping
process in the low-energy limit have the symmetry group
(\ref{symbreakpatt}) \cite{balents98} so that our duality
approach is also relevant to this case.

%%%%%%%%%%%%%%%%%%%%%%%%%%%%%%%%%%%%%%%%%%%%%

\subsection{Low-energy Hamiltonian}

Let us now derive the continuous description of model (\ref{hundmodel}).
The procedure is standard and makes use of the continuum limit
of the lattice fermionic operators $c_{i,l\sigma}$ in terms of
right-and left-moving Dirac
fermions $\Psi_{\ell\sigma \R,\L}$ \cite{bookboso,giamarchi}:
\be
\frac{c^{\vphantom{\dagger}}_{i,\ell\sigma}}{\sqrt{a_0}}
\simeq
\Psi_{\ell\sigma \L}(x)e^{-ik_\f x} +\Psi_{\ell\sigma \R}(x)e^{+ik_\f x},
\ee
with $x=ia_0$.
We now introduce the SU(4)$_1$ currents which are defined from
chiral fermion bilinears:
${\cal J}^A_{\R,\L} = 2 \pi \Psi^{\dagger}_{\ell\sigma \R,\L}
T^A_{\ell\sigma,\ell'\sigma'}\ \Psi_{\ell'\sigma' \R,\L}$,
$T^A$ being SU(4) generators in the fundamental representation which are
normalized according to $\tr(T^AT^B)=\delta^{AB}$.
The set of these 15 generators $T^A$ can be organized
as 3 SU(2)$_{\ind{spin}}$ generators $T_s^{a}$, $a=x,y,z$,
that mix spin indices $\sigma$, 3 SU(2)$_{\ind{orb}}$
generators $T^{a}_t$, $a=x,y,z$, that mix
orbital indices $\ell$, and 9 mixed spin-orbital
generators that act simultaneously in spin and
orbital spaces, $T^{ab}_{st}$, $a,b=x,y,z$.
\be
T_s^a=\frac{1}{2}\,\sigma^a\otimes\idy,
\quad T_t^a=\frac{1}{2}\,\idy\otimes \sigma^a,
\quad T_{st}^{ab} = \frac{1}{2}\,\sigma^a\otimes\sigma^b,
\ee
where $\sigma^a$ are Pauli matrices and
the first (respectively second) matrix in
the tensor product acts on spin (orbital respectively) indices.
Correspondingly, we will introduce the following decomposition
for the 15 components of the SU(4)$_1$ current:
\bea
{\cal J}^a_{s \R(\L)} &=& 2 \pi \Psi^{\dagger}_{\ell\sigma \R (\L)}
\left(T^a_{s}\right)_{\ell\sigma,\ell'\sigma'}
\Psi_{\ell'\sigma' \R (\L)} , \;
{\cal J}^a_{t \R(\L)} = 2 \pi \Psi^{\dagger}_{\ell\sigma \R (\L)}
\left(T^a_{t}\right)_{\ell\sigma,\ell'\sigma'}
\Psi_{\ell'\sigma' \R (\L)} \nonumber
\\
{\cal J}^{ab}_{st \R(\L)} &=& 2 \pi \Psi^{\dagger}_{\ell\sigma \R (\L)}
\left(T^{a b}_{st}\right)_{\ell\sigma,\ell'\sigma'}
\Psi_{\ell'\sigma' \R (\L)} .
\label{su4level1currents}
\eea

For incommensurate filling, the low-energy Hamiltonian
separates into two commuting pieces:
${\cal H} = {\cal H}_c + {\cal H}_s$ ($\left[{\cal H}_c, {\cal H}_s \right] = 0$)
where the charge degrees of freedom are described by a Luttinger Hamiltonian:
\be
{\cal H}_c = \frac{v_c}{2} \left[ \frac{1}{K_c}
\left(\partial_x \Phi_c \right)^2 + K_c
\left(\partial_x \Theta_c \right)^2 \right],
\ee
with the following Luttinger parameters for model (\ref{hundmodel}):
\bea
v_c &=& v_{\f} \left( 1 + a_0 \; \frac{6 U - J_t}{2 \pi v_{\f}} \right)^{1/2}
\\
\nonumber
K_c &=& \left( 1 + a_0 \; \frac{6 U - J_t}{2 \pi v_{\f}} \right)^{-1/2} .
\label{luttparameterhund}
\eea
For generic fillings, no umklapp terms appear and the
charge sector displays metallic properties
in the Luttinger liquid universality class. All non-trivial physics
corresponding to the spin degeneracy is described by the spin part, i.e., ${\cal H}_s$.
Neglecting spin-velocity renormalization terms, the latter Hamiltonian
corresponds to an anisotropic SU(4) model with
marginal current-current interactions:
\begin{equation}
\Cal H_s =
\frac{v_{\f}}{ 20 \pi} \;
\left(\J_{\R}^A \J_{\R}^A + \J_{\L}^A \J_{\L}^A
\right)
+ \; \sum_{\alpha = 1}^{5}
g_{\alpha} \;{\J}_{\R}^A \; d^{\alpha}_{AB} \;{\J}_{\L}^B,
\label{hamiltonianhund}
\end{equation}
where the five matrices $d^{\alpha}$ encode
the symmetry group (\ref{symbreakpatt}) of the low-energy effective Hamiltonian.
Those matrices are diagonal and they read
explicitly (we choose the following ordering
for the 15 SU(4) currents: $T^a_s,T_t^a,T_{st}^{ax},T_{st}^{ay},T_{st}^{az}$):
\bea
d ^{1} &=&  \left(\begin{array}{cccccc}
  \idy_3 & \\
   & 0_{12}
\end{array}\right),
\hspace*{0.5cm} d ^{2} =  \left(\begin{array}{cccccc}
  0_6 & \\
 & \idy_{6} \\
 &  & 0_{3}
\end{array}\right),
\hspace*{0.5cm} d ^{3} =  \left(\begin{array}{cccccc}
  0_{12} & \\
 & \idy_{3}
\end{array}\right),
\nonumber \\
d ^{4} &=&  \left(\begin{array}{cccccc}
  0_3 & \\
 & \idy_{2} \\
 &  & 0_{10}
\end{array}\right),
\hspace*{0.5cm} d ^{5} =  \left(\begin{array}{cccccc}
  0_5 & \\
 & \idy_{1} \\
 &  & 0_{9}
\end{array}\right).
\label{dmatrix}
\eea
In the continuum limit, and at first order in the lattice couplings $(U,J_t,J_H)$,
the coupling constants in Eq. (\ref{hamiltonianhund})
are given by:
\bea
g_1 &=& - \frac{a_0}{8\pi^2} \left( 2 U + J_t - J_H \right),
\hspace*{0.5cm} g_2 =  \frac{a_0}{16\pi^2} \left(- 4 U + 2 J_t + J_H \right),
\hspace*{0.5cm} g_3 = - \frac{a_0}{8\pi^2} \left(2 U +  J_t + J_H \right),
\nonumber \\
g_4 &=&  - \frac{a_0}{16\pi^2} \left(4 U - 2 J_t + 3 J_H \right),
\hspace*{0.5cm} g_5 =  - \frac{a_0}{8\pi^2} \left(2 U - 3 J_t \right) .
\label{contlimithund}
\eea

%%%%%%%%%%%%%%%%%%%%%%%%%%%%%%%%%%%%%%%%%%%%%
%%%%%%%%%%%%%%%%%%%%%%%%%%%%%%%%%%%%%%%%%%%%%
%%%%%%%%%%%%%%%%%%%%%%%%%%%%%%%%%%%%%%%%%%%%%

\subsection{Duality approach}

We are now in position to apply the general duality approach
to the specific model (\ref{hamiltonianhund}) to fully determine the nature
of its spin-gapped phases. The general phase diagram of the lattice
model (\ref{hundmodel}) will then be deduced.

\subsubsection{Duality symmetries}

The Hamiltonian in the spin sector (\ref{hamiltonianhund})
takes the form of an SU(4) anisotropic current-current model
so that we can apply the general result of section \ref{secferm}
for the special $N=4$ case.
In this respect, we find three non-trivial duality symmetries
which exhaust the possible classes for incommensurate filling of the general
classification listed in Table \ref{tabZ2}.

A first one, to be called $\Omega_1$, belongs to
the A\I class of symmetric pairing
with $\frg_\parallel$
$ = \mathfrak{so}$($4$) (see Table \ref{tabZ2}).
It is associated to the following involutive element
of the center of the symmetry
group H: $c_{l \sigma} \to R_{l \sigma, l^{'} \sigma^{'}}
c^\dagger_{l^{'} \sigma^{'}}$, where
the symmetric matrix $R$
reads: $R=-\sigma^y\otimes\sigma^y$ (in a good basis $\tilde c_a$ for
the fermions, this corresponds
to $\tilde c_a \too \tilde c^\dagger_a$).
At the level of the SU(4)$_1$  currents, $\Omega_1$ affects
the following components:
\be
\Omega_1:
\begin{array}{lcll}
\J^{ab}_{st \L} &\rightarrow & - \J^{ ab}_{st \L},& a,b=x,y,z.
\end{array}
\label{om1}
\ee
This transformation is indeed a symmetry of Eq. (\ref{hamiltonianhund})
provided that the couplings are changed as follows: $g_{2} \rightarrow - g_{2}$ and
$g_3\to -g_3$.

A second one, to be called $\Omega_2$,
belongs to type A\II of antisymmetric pairing with
$\fr g_\parallel = \fr{sp}(4)$. It is associated to the following antisymmetric
charge conjugation symmetry:
$c_{l \sigma} \to J_{l \sigma, l^{'} \sigma^{'}}
c^\dagger_{l^{'} \sigma^{'}}$
where
the antisymmetric matrix $J$ (the Sp(4) metric in the $c_{\ell\sigma}$ basis) reads:
$J=i\sigma^y\otimes\sigma^x$. The
SU(4)$_1$ currents affected by $\Omega_2$ are:
\be
\Omega_2:
\left\{
\begin{array}{lcll}
\J^{x}_{t \L} &\rightarrow & - \J^{x}_{t \L} &\\
\J^{y}_{t \L} &\rightarrow & - \J^{y}_{t \L} &\\
\J^{a z}_{st \L} &\rightarrow & - \J^{ a z}_{st \L},& a=x,y,z.
\end{array}
\right.
\label{om2}
\ee
It has the following representation on the couplings:
$g_3\to -g_3$ and $g_4\to -g_4$.

The last non-trivial duality, $\Omega_3$,
belongs to the A\III class with
$\frg_\parallel = \fr s(\fr u(2)\times\fr u(2))
= \fr u(1)\oplus \fr{su}(2)\oplus \fr{su}(2)$,
and is associated with the orbital rotation of angle $\pi$:
$c_{\ell\sigma}\to -i(-)^\ell\,c_{\ell\sigma}$ which is indeed an
involutive symmetry of the bare
Hamiltonian commuting with all other symmetries. It affects the SU(4)$_1$
currents as follows:
\be
\Omega_3:
\left\{
\begin{array}{lcll}
\J^{x}_{t \L} &\rightarrow & - \J^{x}_{t \L} &\\
\J^{y}_{t \L} &\rightarrow & - \J^{y}_{t \L} &\\
\J^{a x}_{st \L} &\rightarrow & - \J^{ a x}_{st \L},& a=x,y,z \\
\J^{a y}_{st \L} &\rightarrow & - \J^{ a y}_{st \L},& a=x,y,z.
\end{array}
\right.
\label{om3}
\ee
This duality has the following action on the couplings:
$g_{2} \rightarrow - g_{2}$ and
$g_4\to -g_4$.

Amongst those non-trivial dualities, two of them, namely $\Omega_1$ and $\Omega_2$,
are \emph{outer} dualities.
It is coherent to christen the fourth duality, namely the trivial
(identity) operation, $\Omega_0$.
Note that the set of the 4 dualities $\{\Omega_a\}_{a=0,1,2,3}$ has the structure
of the Klein four-group, or $\Z_2\times\Z_2$, with the following
multiplication table for non-trivial elements: $\Omega_1\Omega_2=\Omega_3$,
$\Omega_1\Omega_3=\Omega_2$, $\Omega_2\Omega_3=\Omega_1$.

Before we move on to the description of the phase diagram, it is useful to give an
alternative form for the low-energy effective Hamiltonian. Indeed, there is a
possible representation of the unperturbed SU(4)$_1$ model in terms of 6 free
Majorana fermions, which has been widely used in the literature on the 2-leg
ladders
\cite{schulz,aza4spin,lee,controzzi,
totsuka,tsuchiizu05,esslerkonik,bunder}. This Majorana description
of the SU(4)$_1$ WZNW model
is the quantum translation of the classical equivalence between the Lie algebras of
SU(4) and SO(6).
As we will see, this equivalence also allows for a very
convenient representation of the possible dualities, that are realized as SO(6)
dualities.

To this end, one introduces six free massless
Majorana fermions $\xi^a_{\L(\R)}$
 that allow to represent the non-interacting part
of model (\ref{hamiltonianhund}) as
\be
\Cal H_{s0} = -\frac{iv_\f}{2}\sum_{a=1}^6\left( \xi^a_\R\p_x\xi^a_\R
- \xi^a_\L\p_x\xi^a_\L  \right).
\ee
Under SU(2)$_{\ind{spin}}$ rotations, the three Majorana $\xi^{a}$ ($a=1,2,3$)
transform as a spin one, while the three
remaining Majorana $\xi^{a}$ ($a=4,5,6$)
transform as a spin one under orbital rotations of SU(2)$_{\ind{orb}}$.
The SU(4)$_1$ currents can then be written as fermionic bilinears. Explicitly, the
triplet of spin currents reads
$\J^a_{s \L(\R)}=-i\pi\epsilon^{abc}\xi^b_{\L(\R)}\xi^c_{\L(\R)}$,
the triplet of orbital currents reads
$\J^a_{t \L(\R)}=-i\pi\epsilon^{abc}\xi^{b+3}_{\L(\R)}\xi^{c+3}_{\L(\R)}$,
while the remaining 9 mixed spin-orbital  SU(4)$_1$ currents read
$\J^{ab}_{st \L(\R)}=-2i\pi\xi^{a}_{\L(\R)}\xi^{b+3}_{\L(\R)}$
($a,b = 1,2,3$).
The full interacting
Hamiltonian (\ref{hamiltonianhund}) can then be conveniently written as:
\bea
{\Cal H}_s = \Cal H_{s0} + 4\pi^2 \left[
 g_1\sum_{1\leq a < b \leq 3} \kappa^a\kappa^b
+g_2\sum_{1\leq a \leq 3} \kappa^a(\kappa^4 + \kappa^5)
+g_3\sum_{1\leq a \leq 3} \kappa^a\kappa^6
+g_4(\kappa^4 + \kappa^5)\kappa^6
+g_5 \kappa^4  \kappa^5
\right] ,
\eea
where one has introduced the energy density (or ``thermal operator'')
$i\kappa^a = i\xi^a_{\R}\xi^a_{\L}$ of the Ising model associated to each of
the 6 free Majorana fermion theories.

In terms of these fermionic variables, the three non-trivial dualities
(\ref{om1},\ref{om2},\ref{om3}) read:
\bea
\Omega_1: & \xi_{\L}^a\to - \xi_{\L}^a \qquad & a=4,5,6 \nonumber\\
\Omega_2: & \xi_{\L}^6\to - \xi_{\L}^6 \qquad & \label{fermdual}\\
\Omega_3: & \xi_{\L}^a\to - \xi_{\L}^a \qquad & a=4,5 . \nonumber
\eea
In other words, the non-trivial dualities
can be constructed from ``elementary''
KW dualities on the Ising models attached
to each Majorana fermion theory.
This is very peculiar to the $N=4$ case, i.e. two-leg ladders, and
this result does not generalize for $N >4$.
Of course, restrictions on the way they can be
combined come from demanding that they be compatible with H-invariance.
Note that
all of these dualities, once expressed in the fermionic language, become SO(6)
dualities of type BD\I (see Table \ref{tabZ2}). By using the well-known Lie algebra
isomorphisms $\fr{u}(1)=\fr{so}(2)$, $\fr{su}(2)=\fr{so}(3)$,
 $\fr{sp}(4)=\fr{so}(5)$ and
$\fr{so}(4)=\fr{su}(2)\oplus\fr{su}(2)$,
one can easily check that for each duality
$\Omega_1,\Omega_2,\Omega_3$ respectively,
the fermionic representation  (\ref{fermdual})
yields the correct invariant subspace $\fr g_\parallel$, namely
$\fr g_\parallel = \fr{so}(3)\oplus\fr{so}(3), \fr{so}(5), \fr{so}(2)\oplus\fr{so}(4)$
respectively.

\subsubsection{Phase diagram}

According to the analysis of section \ref{secferm}, the knowledge of
the different
allowed dualities $\Omega_a$ gives direct access to the
different symmetry enlarged
phases $\M_a$.
Explicitly, the lattice order
parameters, that will develop
quasi-long-range
order in phase $\M_a$ at incommensurate filling are obtained from the
``fundamental'' CDW order parameter (${\Cal O}^{(CDW)}_0$)
 by acting on one of the two
involved lattice fermions with the
involutive
lattice symmetry associated
to the duality $\Omega_a$ (see Table II).
Using  the dualities (\ref{fermdual}) that we have found,
we can therefore readily conclude that model (\ref{hundmodel}) with
bare symmetry H (\ref{symbreakpatt}) can support
the following symmetry enlarged
phases:

$\bullet$ Two phases $\M_0$ and $\M_3$ with coexisting SP  and
CDW instabilities at wave vector $2k_\f$. Both CDW-SP phases are
distinguished by their properties under the $\Z_2$ exchange of the two chains:
$\M_0$ is even while $\M_3$ is odd.
The explicit lattice order parameters of $\M_{0,3}$ phases and their continuum
descriptions read as follows:
\be
\begin{array}{rlcl}
\M_0:
&  \;
\Cal O_0^{(CDW)} = {\Cal O}_0
= \sum_{\ell \sigma} \Psi^\dagger_{\ell\sigma\L}\Psi_{\ell\sigma\R}
& \longleftarrow&
e^{-i 2 k_\f j a_0} \sum_{\ell \sigma}
c^\dagger_{j,\ell\sigma}c_{j,\ell\sigma}
\\
&  \;
\Cal O_0^{(SP)}
=e^{ik_\f a_0} \sum_{\ell \sigma}
\Psi^\dagger_{\ell\sigma\L}\Psi_{\ell\sigma\R}
& \longleftarrow&
e^{-i 2 k_\f j a_0} \sum_{\ell \sigma}
c^\dagger_{j,\ell\sigma}c_{j+1,\ell\sigma}
\\
&&&
\\
\M_3:
&  \;
\Cal O_\pi^{(CDW)} = {\Cal O}_3
= \sum_{\ell \sigma}
(-)^{\ell+1}\;\Psi^\dagger_{\ell\sigma\L}\Psi_{\ell\sigma\R}
& \longleftarrow&
e^{-i 2 k_\f j a_0}
\sum_{\ell \sigma}
\;(-)^{\ell+1}c^\dagger_{j,\ell\sigma}c_{j,\ell\sigma}
\\
&  \;
\Cal O_\pi^{(SP)}
=e^{ik_\f a_0}\;
\sum_{\ell \sigma}
(-)^{\ell+1}\Psi^\dagger_{\ell\sigma\L}\Psi_{\ell\sigma\R}
& \longleftarrow&
e^{-i 2 k_\f j a_0}
\sum_{\ell \sigma}
\;(-)^{\ell+1}c^\dagger_{j,\ell\sigma}c_{j+1,\ell\sigma}
\end{array}
\label{DiagonalOP}
\ee

$\bullet$ Two phases $\M_1$ and $\M_2$ characterized
by a superconducting pairing
$\Cal O_1$ ($\Cal O_2$ respectively) instability at wave vector $k=0$, are of
BCS type with $s$-
wave symmetry ($d$-wave symmetry
respectively). The two BCS phases, are
distinguished by the symmetry of the pairing operators:
$\Cal O_1$ (BCS$s$) is odd  under chain exchange, while $\Cal O_2$
(BCS$d$) is even.
\be
\begin{array}{rlcl}
\dr{BCS}s =\M_1:
&  \;
\Cal O_s^{(SC)}=\Cal O_1  =
 \Psi^\low_{a\L} R_{ab}\Psi^\low_{b\R}
&\stackrel{\simeq} \longleftarrow&
 c^\low_{j,a}R_{ab}c^\low_{j+1,b}
=
c^\low_{j+1, 1\upa}c^\low_{j,2\down}
  - c^\low_{j+1,1\down}c^\low_{j,2\upa} - (1\oto 2)
\\
\dr{BCS}d =\M_2:
&  \;
\Cal O_d^{(SC)}=\Cal O_2
=\Psi^\low_{a\L} J_{ab}\Psi^\low_{b\R}
& \longleftarrow&
\frac{1}{2} c^\low_{j,a} J_{ab} c_{j,b}^\low
=
c^\low_{j,1\upa,j}c^\low_{j,2\down}
- c^\low_{j,1\down}c^\low_{j,2\upa},
\end{array}
\label{OffdiagonalOP}
\ee
with $a = (\ell, \sigma)$ and $b = (\ell^{'}, \sigma^{'})$.
The BCS instability is accompanied with a generalized spin superfluidity,
that would reveal itself for example in a rotating
system put  on a ring, with a quantization of the
current in a mixed spin-orbital direction.
As it has been described in section IV C, this result stems from
the fact that $\Omega_1$ and $\Omega_2$ are
outer dualities.
Moreover, if the charge sector is gapped by some process (for example,
by considering a suitable commensurate filling away from half filling),
one can also characterize the BCS phases by the
appearance
of coherence of the spin excitations
($\delta$-peak in the zero temperature spin structure factor),
contrarily to the CDW-SP phases which correspond to inner
dualities.

The general analysis of section  \ref{secferm}
also allows for a determination of
the quantum phase transitions between the  SU(4)-symmetry enlarged phases:
there are as many different
transitions as there are non-trivial dualities.
Close to  a quantum phase transition,
the physics of the two-leg ladder spin sector is
captured by a minimal theory with special
self-dual symmetry (which is simply the invariant subspace of the
$\Z_2$ grading $\Omega_a$),
which is maximal in some sense:
quantum phase transitions require in general that the bare
anisotropy be large enough so that DSE
cannot fully, but only partially,
develop. What happens is sketched in the following diagram:
\begin{center}
\drawsymbreak ,
\end{center}
with the bare theory symmetry indicated on the left,
and the RG flows towards low
energy when moving to the right. In generic situations (i.e. far away from
multicritical points), there is a hierarchy of couplings and DSE
develops in several steps (in our case, two steps) as long as the
energy cut-off is reduced.
Close to a quantum phase transition
 this process is stopped at an intermediate stage, and
the low-energy
theory has a an intermediate symmetry that eventually labels the quantum
phase transition.
We list in Table IV the three different quantum phase transitions
occurring in the model.

\begin{table}[h]
\begin{center}
\begin{tabular}{|c|c|c|c|}
\cline{2-4}
\multicolumn{1}{l|}{} & Symmetry &
Order & Critical degrees of freedom  \\ \hline
$\begin{array}{c}\M_0\oto\M_1\\\dr{and}\\ \M_2\oto\M_3
\end{array}$ & SO(3)$\times$SO(3)
& $\begin{array}{c}1\\2\\2\end{array}$ & $\begin{array}{ll}
\emptyset&{\quad\quad} g_1,g_4=g_5>0\\
\dr{SO(3)}_1\quad (c=\frac{3}{2}) &{\quad\quad} g_1g_4=g_1g_5<0 \\
\dr{SO(6)}_1 \quad (c=3) &{\quad\quad} g_1,g_4=g_5<0\end{array}$
\\\hline
$\begin{array}{c}\M_0\oto\M_2\\\dr{and}\\ \M_3\oto\M_1
\end{array}$ & SO(5)$\times\Z_2$
& $\begin{array}{c}2\\2\end{array}$ & $\begin{array}{ll}
\dr{Ising }\;(c=\frac{1}{2}) &{\quad\quad} g_1=g_2=g_5>0\\
\dr{SO(6)}_1 \quad(c=3) &{\quad\quad} g_1=g_2=g_5<0 \end{array}$
\\ \hline
$\begin{array}{c}\M_0\oto\M_3\\\dr{and}\\ \M_2\oto\M_1
\end{array}$ & SO(4)$\times$SO(2)
& $\begin{array}{c}2\\2\end{array}$ & $\begin{array}{ll}
\dr{L.L. }\;(c=1) &{\quad\quad} g_1=g_3>0\\
\dr{SO(6)}_1 \quad(c=3) &{\quad\quad} g_1=g_3<0 \end{array}$
\\ \hline
\end{tabular}
\end{center}
\caption{Summary of the different possible quantum
phase transitions for the generalized two-leg electronic
ladder (\ref{hamiltonianhund}) for incommensurate filling.}
\end{table}

The next question
is whether
those four phases are actually realized
in the lattice model (\ref{hundmodel}), and
what are the values of the microscopic couplings that will yield each
of those phases.
To answer this question,
one needs to perform, in a weak-coupling approach, a RG
analysis of the continuous effective
theory (\ref{hamiltonianhund}).
The two-loop $\beta$-function can be obtained for
example
in a point splitting regularization \cite{leclair} to yield:
\bea
\dot{g_1} &=& (1-g_1)(g_1^2+2g_2^2+g_3^2)
\nonumber\\
\dot{g_2} &=& g_2(2g_1+g_5)+g_3g_4 - \frac{g_2}{2}(2g_1^2+3g_2^2+g_3^2+g_4^2+g_5^2)
\nonumber\\
\dot{g_3} &=& 2g_1g_3 +2g_2g_4 - g_3(g_1^2+g_2^2+g_3^2+g_4^2)
\label{RGeq}
\\
\dot{g_4} &=& g_4g_5 +3g_2g_3 - \frac{g_4}{2}(3g_2^2+3g_3^2+g_4^2+g_5^2)
\nonumber\\
\dot{g_5} &=& (1-g_5)(3g_2^2+g_4^2) \nonumber,
\eea
where $\dot g=\frac{dg}{d\ln(a/a_0)}$,
with $a_0^{-1}$ the UV cutoff (which is of
the order of the inverse of the lattice spacing) and $a^{-1}$ a
running RG scale, and the couplings
in Eqs. (\ref{RGeq}) have been rescaled
according to
$g_a\to 2\pi g_a/ v_\f$.

\begin{figure}[h]
\begin{center}
\vspace{0.5cm}
\includegraphics[angle=0,width=0.7\linewidth]{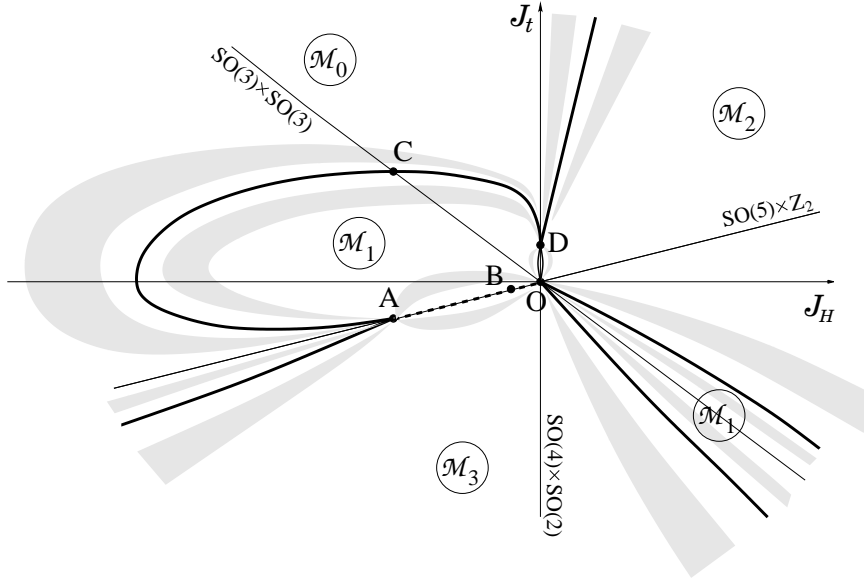}
\caption{Phase diagram of model (\ref{hundmodel}). The Coulombic
repulsion $U=0.01 t$
is fixed, while Hund coupling $J_H$ and orbital anisotropy $J_t$
are varied. Bold lines
indicate the location of the transitions between the
different symmetry enlarged phases $\M_a$
whose nature is discussed in the text. The small pocket lying
between points $O$ and $D$
is of type $\M_3$. The dashed line extending between points
$O$ and $A$ is fully critical,
with central charge $c=3$.
Thin rays correspond to fine-tuned bare theories
with a larger, exact, bare symmetry H' with H=SU(2)$\times$U(1)$\times \Z_2\subsetneq$ H'$\subsetneq$ SU(4)=G. 
On those lines, the effective theory has symmetry H' at any stage of the RG flow, but flows
at low energies to a symmetry enlarged phase with symmetry SU(4) in the sense of section \ref{dualDSE}. 
 On point $O$, the
bare theory is maximally symmetric, with a SU(4) invariance.
On point $A=(-8U,-2U)$, the continuous
theory enjoys an enlarged \emph{twisted} SU(4) symmetry $\Omega_2$(SU(4)).
Self-dual manifolds intersect the $(J_H,J_T)$ plane at the discrete points
$B=(-\frac{8}{5}U,-\frac{2}{5}U)$, $C=(-8U,6U)$ and $D=(0,2U)$,
with $\Omega_1(C)=C$, $\Omega_2(B)=B$ and
$\Omega_3(D)=D$.}
\label{phaseSU2U1Z2}
\end{center}
\end{figure}

One observes that the RG equations are invariant under the
duality action on the couplings,
as it should be (in fact, since dualities are exact
symmetries of the continuous model
(\ref{hamiltonianhund}), the all-order
$\beta$-function is invariant). Numerical integration of the
RG
equations yield the phase diagram presented in figure \ref{phaseSU2U1Z2}.
The four possible symmetry enlarged phases
$\M_0$ to $\M_3$ are reached within
this model for repulsive Coulombic
interaction $U$ (see figure \ref{phaseSU2U1Z2}).

DSE ensures that all bare theories sitting between the transition
lines
(bold lines) has a low-energy physics described by an
adiabatic deformation of the
corresponding
representative ($\Omega_a$-twisted) SU(4) GN model. Closer to transition lines,
a cross-over regime
occurs (gray areas). Even closer to them, one
finds again a universal regime described by the
quantum phase transition.
Note that the curvature of the phase transition lines visible in figure
\ref{phaseSU2U1Z2}, is \emph{not} a two-loop effect, but rather stems
 from the non-linearity of the flow equation that are  already present
at one-loop order. One notices a quite
complex topography of the phase diagram, with for example
an $s$-wave superconducting pocket for ferromagnetic Hund coupling.

\subsection{Allowing inter-leg hopping}

In this section we show how the previous discussion can also be applied to the
case of a two-leg ladder with legs connected by transversal hopping $t_\perp$,
with a SU(2) spin symmetry, and a
$\Z_2$ symmetry that exchanges the two
legs, but no U(1) orbital symmetry. This model
has been examined in several works, with the conclusion
that there were four phases
at incommensurate
fillings \cite{balents98,furusaki,wu04,tsuchiizu05,controzzi}.
We will see that this model is in fact in closely related to the
SU(2)$\times$U(1)$\times\Z_2$ model studied in the preceding
subsections: it is simply obtained therefrom by acting with a duality.

Introducing the bonding and anti-bonding modes that diagonalize the kinetic
term, and expanding those modes around the corresponding
two Fermi points $k_{\f 1}$ and
$k_{\f 2}$, one obtains fermionic
fields $\tilde\Psi_{\ell\sigma\P}$, $\sigma=\upa,\down$,
$\ell=1,2$, $\P=\L,\R$.
The Fermi wave vectors satisfy
$k_{\f+}\equiv k_{\f 1}+ k_{\f 2}=\frac{\pi n}{ a_0}$, with $n$ the
electronic density, while the difference
$k_{\f-}\equiv k_{\f 1}- k_{\f 2}$ is a function of $t_\perp$
and $n$. At fillings and $t_\perp$ such
that $k_{\f-}$ is incommensurate,  it is
easy to see that for an arbitrary $2N$-fermion term
$\prod_{i=1}^N\tilde\Psi^\dagger_{\ell_i\sigma_i\P_i}
\tilde\Psi_{\ell'_i\sigma'_i\P'_i}$
to conserve lattice momentum, it has to conserve \emph{separately} the
combinations $\rho_+=N_{1R}+N_{2L}$ and $\rho_-=N_{2R}+N_{1L}$, where
$N_{\ell\P}=\sum_{\sigma}\tilde
\Psi^\dagger_{\ell\sigma\P}\tilde\Psi_{\ell\sigma\P}$ is
the total number of chiral fermions in mode $\ell$.
Therefore, on top of conserving the
SU(2) spin and the total number of
electron $\rho_++\rho_-$, the Hamiltonian has
to conserve the
combination $\Cal I_t=\frac{1}{2}\left(\rho_+-\rho_-\right)
=\J_{t\R}^z-\J_{t\L}^z$.
This operator is nothing but the
``orbital current'', i.e. the space-like component of
the  Noether current associated with U(1) orbital symmetry.
It results that the model
generically enjoys in fact a larger symmetry
SU(2)$\times\widetilde {\mbox{ U(1) }}\times\Z_2$,
where the twisted orbital symmetry $\widetilde {\mbox{ U(1) }}$
is generated by $\Cal I_t$.

Therefore, in the continuum limit one recovers \emph{exactly}
the model studied
in the preceding subsections, provided one performs a duality $\bar \Omega$
that changes the sign of the following component of the currents
\footnote{Our choice is dictated by the following constraints: $\bar\Omega$
has to preserve the SU(4) Lie algebra (it is an automorphism), it has to leave
$\J^a_s$ invariant, and has to change the sign of $\J^z_t$. There is in fact a
whole family of solutions,
 that are labelled by an angle $\alpha$. Of course, our conclusions do
 not depend on this choice.}:
\be
\bar\Omega:
\left\{
\begin{array}{lcll}
\J^{x}_{t \L} &\rightarrow & - \J^{x}_{t \L} &\\
\J^{z}_{t \L} &\rightarrow & - \J^{z}_{t \L} &\\
\J^{a y}_{st \L} &\rightarrow & - \J^{ a y}_{st \L},& a=x,y,z.
\end{array}
\right.
\label{ombar}
\ee
One thus immediately deduces that the maximal
number of symmetry enlarged phases is four.
Since (\ref{ombar}) corresponds to an outer
automorphism of
$\fr{su}(4)$, one readily knows that pairing phases
are mapped onto density-wave phases, and
vice versa. Note that in terms of the six
Majorana fermions $\xi^a$, this duality bears the very
simple form: $\xi^5_\L\too-\xi^5_\L$.
Using the representation of $\bar\Omega$ on the original
bonding and anti-bonding fermions:
\be
\tilde\Psi_{\ell\upa\L}\too  + \tilde \Psi^\dagger_{\ell\down\L}
\quad,\qquad
\tilde\Psi_{\ell\down\L}\too - \tilde \Psi^\dagger_{\ell\upa\L},
\ee
one can readily identify the four phases $\tilde\M_a=\bar\Omega(\M_a)$ of the $t_\perp$-
ladder, with order parameters $\widetilde{\Cal O}_a=\bar\Omega(\Cal O_a)$
that read explicitly:
\bea
\widetilde{\Cal O}_0 &=&
\sum_\ell\left( \tilde\Psi_{\ell\down\L} \tilde\Psi_{\ell \upa\R}
- \tilde\Psi_{\ell\upa\L} \tilde\Psi_{\ell\down\R}
\right)
\nl
\widetilde{\Cal O}_1 &=&
 \sum_{\sigma}\left(
\tilde\Psi^\dagger_{1\sigma\L}
\tilde	\Psi_{2\sigma\R}^\low -
\tilde\Psi^\dagger_{2\sigma\L} \tilde	\Psi_{1\sigma\R}^\low
\right)                    
\nl
\widetilde{\Cal O}_2 &=& \sum_{\sigma}\left(
\tilde\Psi^\dagger_{1\sigma\L}
\tilde	\Psi_{2\sigma\R}^\low +
\tilde\Psi^\dagger_{2\sigma\L} \tilde	\Psi_{1\sigma\R}^\low
\right)
\nl
\widetilde{\Cal O}_3 &=&
\sum_\ell(-)^{\ell+1}\left( \tilde\Psi_{\ell\down\L} \tilde\Psi_{\ell \upa\R}
- \tilde\Psi_{\ell\upa\L} \tilde\Psi_{\ell\down\R}
\right) .
\nonumber
\eea
These phases correspond to the four phases found
in generalized two-leg ladders with interchain hopping
for incommensurate filling:
coexistence of CDW and SP phases,
d-wave and s-wave superconducting phases (DSC and SSC respectively)
and coexistence of time-reversal breaking phases
like d-density wave (DDW) and diagonal current (DC) phases
(see for instance Ref. \cite{wu04}).
The explicit connection, together with
lattice order parameter in terms of the
original fermions $c_{1}$ and $c_{2}$ on the two legs
of the ladder, is given by:
\be
\begin{array}{rlcl}
\dr{SSC}=\tilde\M_0:
&  \;
\Delta_s =-  \widetilde{\Cal O}_0
& \longleftarrow&
c_{j,1\upa}c_{j,1\down}
+ c_{j,2\upa}c_{j,2\down}
\\
\dr{DSC}=\tilde\M_3:
& \;
\Delta_d =- \widetilde{\Cal O}_3
&\longleftarrow&
c_{j,1\upa}c_{j,2\down}
- c_{j,1\down}c_{j,2\upa}
\\
\dr{CDW+SP}=\tilde\M_2:
&\;
\Cal O^{CDW} = 2 \re (e^{ik_{\f+} x}\;\widetilde{\Cal O}_2)
&\longleftarrow&
 \sum_{\ell,\sigma}
(-)^{\ell+1}c^\dagger_{j,\ell\sigma}c_{j,\ell\sigma}
\\
&\;
\Cal O^{SP} = 4 \cos(k_{\f-}\frac{a_0}{2}) \re (e^{ ik_{\f+}(x+\frac{a_0}{2})}\; \widetilde{\Cal O}_2)
&\longleftarrow&
\sum_{\ell,\sigma} (-)^{\ell+1}c^\dagger_{j,\ell\sigma}c_{j+1,\ell\sigma}
 + \hc
\\
\dr{DDW+DC}=\tilde\M_1:
&\;
\Cal O^{DDW} = - 2  \im (e^{ik_{\f+} x}\; \widetilde{\Cal O}_1 )
&\longleftarrow&
i\sum_\sigma c^\dagger_{j,2\sigma}c_{j,1\sigma} + \hc
\\
&\;
\Cal O^{DC} =  4  \sin(k_{\f-}\frac{a_0}{2})  
\re (e^{ ik_{\f+}(x+\frac{a_0}{2})}\; \widetilde{\Cal O}_1)
&\longleftarrow&
i\sum_{\ell\sigma}(-)^{\ell+1} \;c^\dagger_{j,\ell\sigma}  c_{j+1,\ell\sigma}
 + \hc 
\end{array}
\nonumber
\ee

We thus see that the duality approach
allows to capture the four phases
of the generalized
two-leg ladder with interchain
hopping at incommensurate filling.
In addition, the nature of the different quantum phase transitions between
these phases is still determined by Table IV.
Finally, our approach leads us to conclude that
more exotic phases can only be found
by (i) considering theories with large anisotropic
bare couplings, i.e. theories that flow,
in the IR limit,
close to the quantum phase transitions described previously, or by (ii)
breaking the remaining bare
$\dr{SU(2)}_{\ind{spin}}\times\Z_2$ symmetry.

%%%%%%%%%%%%%%%%%%%%%%%%%%%%%%%%%%%%%%%%%%%%%
%%%%%%%%%%%%%%%%%%%%%%%%%%%%%%%%%%%%%%%%%%%%%
%%%%%%%%%%%%%%%%%%%%%%%%%%%%%%%%%%%%%%%%%%%%%
\section{Conclusions}

In this paper, we have developed a general non-perturbative
approach to describe
spin-gapped phases of
weakly interacting one-dimensional degenerate fermions.
In the continuum limit, the low-energy properties of
these systems are described by a WZNW CFT
perturbed by marginal relevant current-current interaction with
H invariance.
At the heart of the analysis is the existence, in
this general class of model, of emergent duality
symmetries which enable one to relate different competing
orders between themselves and to shed light on
the nature of the zero-temperature quantum phase transitions.
In particular, we have shown that these dualities can be classified
and depend on the algebraic properties of the problem:
the physical symmetry group H of the system
and the maximal continuous symmetry group
of the interaction G.
For $N$-component degenerate fermions, this maximal
symmetry group is G = SU($N$) or SO($2N$)
away from half-filling and at half-filling respectively.
The duality symmetries can be identified as the involutions
which belong
to the center group $\Cal C\big(\mbox{H}\big)_{\ind{inv}}$
of the symmetry group H of the problem
so that for each duality,
there is an involutive $\Z_2$ lattice symmetry
that gives rise to it.
Alternatively, based on the DSE phenomenon, the duality symmetries
are also in one-to-one correspondence with
the different symmetry enlarged phases that can be supported by the generic
fermionic model with marginal current-current interactions.
Those symmetry enlarged phases are fully gapped phases, and all meet at the
multicritical point, that is obtained by fine-tuning all interactions to zero.
Moreover, and
this should not be a surprise since we are dealing with one-dimensional models,
those
massive phases do not break spontaneously any continuous symmetries. Hence, our
approach allows to draw a picture of a quite large class of possible
generalized spin-liquid phases supported by coupled fermionic chains in one
dimension, at least
those that develop close to the multicritical (non-interacting) point.

In the course of our study, there naturally appears a strong distinction
between two
kinds of gapped spin-liquid phases for incommensurate filling
according to whether the duality they are
associated
to is an  \emph{inner} and \emph{outer} automorphism. This quite mathematical
distinction turns out to have important physical signatures.
``Inner'' phases, appear to
display ``conventional'' properties, in the following
sense: the ordering is of
charge-density type, with a lattice order parameter of the form
$c^\dagger_{j,a}M_{ab}c_{j,b}$.
On the other hand, ``outer'' phases display off-diagonal order,
with the development of
quasi-long-range correlations for pairing operators
$c^\dagger_{j,a}M_{ab}c^\dagger_{j,b}$.
This superconducting instability is accompanied by spin superfluidity:
the low-energy collective modes carry spin
currents. In spite of the gap, the system has non-vanishing
susceptibility in some
spin directions, that results for example in a quantization of the spin current
for a rotating system put on a ring.
 All these ``unconventional'' properties can be ultimately connected to the
spontaneous breaking of a discrete symmetry of the bare theory:
the discrete phase
redefinition of the fermions (\ref{defZN}), which it is tempting to
connect to the usual breaking
of U(1)
gauge invariance in a superconducting state in higher dimensions.

We can also address the issue of quantum phase transitions
between the different
phases, by means of ``minimal theories'' interpolating between them. The nature
of the phase transition between the two symmetry enlarged
phases $\M_{\Omega_1}$ and
$\M_{\Omega_2}$ (associated to dualities $\Omega_1$ and $\Omega_2$) depends
only on the duality $\Omega=\Omega_1\Omega_2$, so that the same set of
dualities can be used to label those minimal models.
The self-dual manifold of these minimal models exhibit in general criticality,
that captures the nature of the quantum phase
transition between the phases.
We have shown in the specific example of
two-leg electronic ladders at incommensurate
filling how this approach allows
for an immediate determination of the possible symmetry enlarged phases on the
basis of the analysis of the bare
symmetry group (of the continuous theory) and
the determination of the different quantum phase transitions.

As perspective,
it would be interesting to study the interplay of symmetry enlargement with
doping -- i.e. how commensurability generically affect the
general picture of the phase diagram.
In this respect, the half-filled case is very special
since there is no spin-charge
separation and
the charge degrees of freedom cannot
be disentangled from the
spin ones due to an umklapp process.
As it has been discussed in
this paper, the relevant maximal symmetry group
for $N$ fermionic species at half-filling is G = SO($2N$).
The detail of the analysis of the dualities for this case
turns out to be
more complicate than for the incommensurate case, i.e. with G = SU($N$),
and will be investigated elsewhere
\footnote{An exception occurs for $N=4$, the 2-leg electronic ladder, for which the group 
identity SU(4)$\sim$SO(6) allows to factorize the SO(8) dualities into a charge part and a 
spin part. One can then show that with respect to the incommensurate case, a 
\emph{doubling} of the number of allowed dualities occurs, yielding 8 different phases for 
the example studied in section \ref{secexample}. It is important to realize that the triality 
-- a remarkable, non-involutive automorphism of SO(8) -- that was invoked in 
Ref.\onlinecite{balents98} 
in the study of the half-filled 2-leg electronic ladder has nothing to do with the dualities 
studied here. Rather, it is connected to the identity SU(4)$\sim$SO(6), and  helps 
understanding \emph{why} $N=4$ is so particular.
We shall come back to this in a forthcoming publication.}.
Dualities of the kind studied in this work can
have further applications in other contexts.
For example, in degenerate quantum impurity problems
where a localized spin is coupled to
electronic spin currents, they can be used
to relate different IR boundary fixed points.
More generally, the duality symmetries can be viewed as automorphisms
of the fusion algebra of a CFT. We focused here on WZNW models perturbed
marginally by current-current interactions.
Generalized dualities could be used to shed light on
the phase diagram of other CFT's perturbed by relevant operators.
In this respect, this case will be useful to investigate
the possible classification of 1D spin-liquid phases
where the spin gap is opened by a strongly relevant perturbation.
We hope to come back to these issues elsewhere.

%%%%%%%%%%%%%%%%%%%%%%%%%%%%%%%%%%%%%%%%%%%%%
%%%%%%%%%%%%%%%%%%%%%%%%%%%%%%%%%%%%%%%%%%%%%
%%%%%%%%%%%%%%%%%%%%%%%%%%%%%%%%%%%%%%%%%%%%%

%%%%%%%%%%%%%%%%%%%%%%%%%%%%%%%%%%%%%%%%%%%%%%%
%%%%%%%%%%
\begin{acknowledgements}
The authors are grateful to
F. H. L. Essler, R. Konik, H. C. Lee, T. Momoi, A. A. Nersesyan,
K. Totsuka, and A. M. Tsvelik for useful discussions.
\end{acknowledgements}

%%%%%%%%%%%%%%%%%%%%%%%%%%%%%%%%%%%%%%%%%%%%%%%
%%%%%%%%%%%%%%%

\appendix

\section{Characterization of dualities}
\label{app-proofdual}

In this Appendix, we first characterize the
set of linear transformations of the currents
that leave the Hamiltonian (\ref{hamiltonianref}) globally invariant,
by (i) showing that these transformations can
be decomposed in the product of diagonal
and chiral transformations and by (ii)
showing that the chiral transformations,
that correspond to dualities, are given
by Eq. (\ref{dualset}), i.e.~they have to belong
to the involutive part of the center
of the bare symmetry group H. Then we show
that for a given model, the set of dualities
coincides with the different possible inequivalent symmetry enlargements.
We recall that
$\du$ denotes the set of allowed dualities, $D$ is the vector space
spanned by the matrices $d^\alpha$.
In particular, $D$ is
the set of real (from hermiticity)
symmetric (from parity) matrices that
commute with the action of H. From
renormalizability, $D$ is closed under
anticommutation: $\acom{D}{D}\subset D$.

\subsection{Dualities are the involutive part of the center of H}
\label{app-dual-Cinv}

We investigate here the general case of
a transformation $\Omega$, that acts
both on the currents and the coupling constants,
such as to leave Hamiltonian (\ref{hamiltonianref})
globally invariant. $\Omega$ acts linearly
 on the currents, and one can write its action as
$\omega_{\L}\times\omega_{\R}$ as in section III:
\be
\tJ_{\L(\R)}^A=\big(\omega_{\L(\R)}\big)^{AB}\; \J_{\L(\R)}^B,
\label{defOmega}
\ee
while it changes the coupling
constants as $g_\alpha\too\tilde g_\alpha=\Omega(g_\alpha)$.
From the invariance of the current
OPE's (\ref{opecurrent}), it follows that
both $\omega_{\L}$ and $\omega_{\R}$ must
be automorphisms of G. One useful remark
is that G-automorphisms automatically
belong to O(dim(G)), i.e. the
matrices $\omega_{\L(\R)}$ are orthogonal.
The covariance of the interacting part of
the Hamiltonian (\ref{hamiltonianref}) leads to an identity that generalizes
Eq. (\ref{faldual}):
\be
g_\alpha \,d^\alpha=\tilde g_\alpha \, ^t\omega_{\R} d^\alpha\omega_{\L},
\label{appdu1}
\ee
where $^t\omega_{\R}$ denotes the transpose of the matrix $\omega_{\R}$.
One starts by decomposing $\omega_{\L}\times\omega_{\R}
= (\omega_{\R}\times\omega_{\R}) \circ  (\omega\times\idy)$,
with $\omega=\omega_{\L}\omega_{\R}^{-1}$,  as the product
of a diagonal and a chiral part. Considering the
special isotropic ray $g_\alpha=g^0_\alpha$,
with $g_\alpha^0 d^\alpha = \idy$, and
multiplying (\ref{appdu1}) by $\omega_{\R}$ on
the left, and by $^t\omega_{\L}$ on the right, one
deduces that the matrix $\omega^{-1}$ belongs to
D: $\omega^{-1}=f_\alpha\,d^\alpha$,
with $f_\alpha=\Omega(g_\alpha^{0})$. It results
that $\omega^{-1}$ is a symmetric matrix;
but $\omega$ being also orthogonal, one
has $\omega^{-1}=\,^t\!\omega=\omega$. We deduce thus
\be
\omega\in \mbox{D}, \mbox{ and } \omega^2 = 1.
\label{step1}
\ee

Now, we want to show that the diagonal
rotation $\omega_{\R}\times\omega_{\R}$ alone
leaves the Hamiltonian (\ref{hamiltonianref}) globally invariant.
This is equivalent
to showing that $^t \omega_{\R}\,d^\alpha\,\omega_{\R}\in
\mbox{D}$ for all $\alpha$. Multiplying
relation (\ref{appdu1}) by $\omega_{\R}$ on the
left and by $\omega_{\R}^{-1}$ on the right, one
gets the following relation, valid
for any couplings $g_\alpha$:
\begin{equation}
 g_\alpha\;\omega_{\R} d^\alpha \;{}^t\omega_{\R}
=\tilde g_\alpha d^\alpha\omega .
\label{dualcondapp}
\end{equation}
Transposing
this relation, and recalling that $\omega$
and $d^\alpha$ are symmetric, we deduce
that $\omega$ commutes with $d^\alpha$ for
all $\alpha$ (for this, one chooses
couplings $g_\beta=\Omega^{-1}(\delta_{\alpha\beta})$,
so that $\tilde g_\beta=\delta_{\alpha\beta}$). It results
that $d^\alpha\omega=\frac{1}{2}\{d^\alpha,\omega\}\in\mbox{D}$ since
$\acom{D}{D}\subset D$.  We deduce then from Eq. (\ref{dualcondapp})
the announced result:
\be
^t \omega_{\R}\,d^\alpha\,\omega_{\R}\in \mbox{D},
\ee
i.e.~the diagonal rotation $\omega_{\R}\times\omega_{\R}$
leaves the Hamiltonian (\ref{hamiltonianref}) globally invariant.
This proves our
claim that the general transformation
$\Omega$ given by Eq. (\ref{defOmega}) can be
decomposed into the product of a diagonal rotation,
that corresponds to a \emph{global} change of
 basis, and of a transformation $\omega$
affecting only one chirality sector, both of
these transformations separately leaving
the Hamiltonian (\ref{hamiltonianref}) globally invariant.

We now proceed to characterize the set of
chiral transformations $\omega$, that it is
 legitimate to term dualities in view of their
involutive character. To study the dualities, we
therefore set $\omega_{\R}=\idy$ in the following,
with no loss of generality, so that $\Omega$ is
a transformation of the left currents
and of the couplings, that satisfies:
\be
g_\alpha d^\alpha = \tilde g_\alpha \; d^\alpha\omega .
\label{appdu2}
\ee
We already know that $\omega$ belongs to
D. Moreover, from Eq. (\ref{appdu2}),
fixing $\alpha$ and choosing $g_\beta=\Omega(\delta_{\alpha\beta})$
(i.e. ${\tilde g}_\beta = \delta_{\alpha\beta}$),
one gets $d^\alpha\omega=\Omega(\delta_{\alpha\beta})d^\beta
\in D$. Transposing this relation, and
recalling that $d^\beta$ and $\omega$ are
symmetric, we deduce that $\com{\omega}{d^\alpha}=0$,
or $\omega d^\alpha\omega=d^\alpha$, $\forall \alpha$.
It results that $\omega$ is an element of
the physical symmetry group H.  Since $\omega\in D$
and $\com{H}{D}=0$, $\omega$ is in the center of H. Thus,
one has proved that $\du\subset\Cal C(\mbox H)\big|_{\ind{inv.}}$.

Reciprocally, let us take
some $\omega\in\Cal C(\mbox H)\big|_{\ind{inv.}}$.
Being involutive and orthogonal, $\omega$ has to
 be a real symmetric matrix. Furthermore,
it commutes with H. This is enough to
ensure that $\omega\in D$. To show that $\omega$ is
a duality, it suffices to prove
that Eq. (\ref{appdu2}) holds. This is easily
done: $d^\alpha\omega=\frac{1}{2}\acom{\omega}{d^\alpha}
\in D$.

This ends the proof of Eq. (\ref{dualset}).

\subsection{Dualities correspond to the different possible DSE's}
\label{app-dual-dse}

It is not difficult to realize that
each duality $\Omega\in\du$ defines a different DSE.
Considering the G-isotropic model $\M_0$
with interacting part of the Hamiltonian
$\Cal H_{\ind{int}}^0 = g\;\J^A_{\R}\J^A_{\L}$
and acting on it
with a duality $\Omega\in\du$, one gets the
model $\M_\Omega$ with interacting
part $\Cal H_{\ind{int}}^\Omega=g\;
\J_{\R}^A\Omega(\J^A_{\L}) = g\; \J^A_{\R}\tJ^A_{\L}$.
Since $\Omega$ is an automorphism of G ($\Omega \in \mbox{Aut(G)}$),
the currents $\tJ^A_{\L}=\Omega(\J^A_{\L})$ are
the generators of the (twisted) group $\tilde{\mbox{G}}_{\L}$.
$\M_\Omega$ is thus obviously invariant
under the group $\big(\tilde{\mbox{G}}_{\L}
\times\mbox{G}_{\R}\big)_{\ind{diag}}$ generated
by $\J^A_{\R}+\tJ^A_{\L}$.

Reciprocally, let us suppose there exists
some global invariance group $\tilde{\mbox{G}}$,
isomorphic to G, that leaves invariant a
model of the form (\ref{hamiltonianref}),
with (global) generators $\tilde{Q}^A$ linearly
related to the original ones in each
chirality sector, $\tilde{Q}^A = \int dx\,
\big( \tJ_{\L}^A + \tJ_{\R}^A\big)$,
with $\tJ^A_{\L(\R)}=\omega_{\L(\R)}^{AB}\J^B_{\L(\R)}$.
Then, considering the non-interacting part
of the Hamiltonian, it follows
that $\omega_{\L(\R)}$ must belong to Aut(G);
considering the interacting part, which must be
of the form $g\,\tJ^A_{\R}\tJ^A_{\L}$
from $\tilde{\mbox{G}}$-invariance, leads to
the conclusion that $\omega\equiv\omega_{\R}^{-1}
\omega_{\L}\in D$. Since $\omega$ belongs
to $D\cap\mbox{Aut(G)}$, following the same line of reasoning as in
\ref{app-dual-Cinv}, we conclude that
$\omega\in \Cal C(\mbox H)\big|_{\ind{inv.}} = \du$. Thus,
up to a global rotation $\omega_{\R}\times\omega_{\R}$
(affecting identically both chirality sectors),
 the generators $\tilde{Q}^A$ are nothing but
the dual generators $\int dx\,\big(\J_{\R}^A+\omega^{AB}\J^B_{\L}\big)$.
This completes the proof.

Hence, it results that the set of dualities identifies with
the different possible DSE patterns, i.e. they correspond to the
different possibilities to glue together the two chiral invariance
groups G$_{\L}$ and G$_{\R}$ in a way consistent with H-invariance.

%%%%%%%%%%%%%%%%%%%%%%%%%%%%%%%%%%%%%%%%%%%%

\end{document}